\documentclass[12pt]{article}
\usepackage{graphicx}
\usepackage{epsfig}
\usepackage{enumerate}
\usepackage{amsmath}

\newcommand{\bea}{\begin{eqnarray}} \newcommand{\eea}{\end{eqnarray}}
\newcommand{\attr}{0}
\newcommand{\half}{\frac{1}{2}}
\newcommand{\refeq}[1]{\stackrel{(\ref{#1})}{=}}
\newcommand{\p}{\partial}
\newcommand{\F}{{\cal E}}

\begin{document}

\setlength{\unitlength}{1mm}

\thispagestyle{empty} 

\vspace*{3cm}

\begin{center}
  {\bf \Large Moduli and (un)attractor black hole thermodynamics}\\
  \vspace*{2cm}

  {\bf Dumitru Astefanesei}\footnote{E-mail: {\tt dastef@mri.ernet.in}}
  {\bf Kevin Goldstein}\footnote{E-mail: {\tt kevin@theory.tifr.res.in}}
  {\bf and Swapna Mahapatra}\footnote{E-mail: {\tt swapna@iopb.res.in}; }

  \vspace*{0.2cm}

  {\it $^1$\sl Harish-Chandra Research Institute}\\
  {\it \sl Allahabad, 211 019, INDIA}\\[.5em]
  {\it $^2$\sl Tata Institute of Fundamental Research, Homi Bhabha Road,}\\
  {\it \sl  Mumbai 400 005, INDIA}\\[.5em]
  {\it $^{3}$\sl  Physics Department, Utkal University, Bhubaneswar 751 004, INDIA}
    \vspace*{3mm } \\[.5em]

  \vspace{2cm} {\bf ABSTRACT}
   \end{center}

   We investigate four-dimensional spherically symmetric black hole solutions in
  gravity theories with massless, neutral scalars non-minimally coupled to gauge fields.
  In the non-extremal case, we explicitly show that, under the variation of the moduli,
  the scalar charges appear in the first law of black hole thermodynamics. In the extremal
  limit, the near horizon geometry is $AdS_2\times S^2$ and the entropy does not
  depend on the values of moduli at infinity. We discuss the attractor behaviour by using
  Sen's entropy function formalism as well as the effective potential approach and their
  relation with the results previously obtained through special geometry method.  We also
  argue that the attractor mechanism is at the basis of the matching between the
  microscopic and macroscopic entropies for the extremal non-BPS Kaluza-Klein black hole.

\vfill \setcounter{page}{0} \setcounter{footnote}{0} \newpage

\newcommand{\csch}{\mathrm{csch}\,}

\section{Introduction}

One of the most successful applications of string theory, as a theory of quantum gravity,
has been the study of black holes and their entropy.  String theory provides a microscopic
description of a special class of black holes \cite{Strominger:1996sh, Maldacena:1996ky}.
Essential to the embedding of charged black holes in string theory is the notion of
compactification on some internal compact manifold.  A fundamental element of these
constructions is that all of the stringy constituents are wrapped around some non-trivial
cycles of the internal space so that the final configuration appears as point-like in the
lower dimensional space.  A sufficiently heavy compactified wrapped object will
effectively give rise to a lower dimensional space containing a horizon, or in other
words, a black hole.  Roughly speaking, the microscopic picture of black holes in string
theory is based on the (tiny) internal manifold where the extended objects are trapped.
The geometry of internal manifold is parametrised by certain $moduli$.  These moduli will
appear as fields in the lower dimensional effective field theory.

It is well-known that the expectation value of the dilaton controls the string coupling
constant $g_s=e^{\left< \phi \right>}$. This tells us that the strength
of the interaction is determined $dynamically$ via the vacuum expectation value (vev) of a
scalar field in the string spectrum.  In fact, even the constants which appear upon
compactification are vevs of certain massless scalar fields (referred to as $moduli$
fields) and are determined dynamically by the choice of the vacuum (i.e., the choice of a
consistent string background).

In this paper we consider static $4$-dimensional charged black hole solutions in gravity
theories with $U(1)$ gauge fields and neutral massless
scalars.\footnote{While massless scalars are unnatural in a generic non-supersymmetric
  theory they are at least technically natural with ${\cal N}=1$ supersymmetry.  For this
  reason, our results are best understood in the context of non-supersymmetric solutions
  of theories with ${\cal N}\geq1$.}
The general Lagrangian we consider includes the bosonic part of ${\cal N} = 1$ or $2$
supergravities for particular values of the couplings. We refer to the scalar fields as
moduli even though they do not necessarily characterize the geometry of an internal space
because they still determine the $U(1)$ couplings. The moduli have a non-trivial radial
dependence and hence the properties of these black holes depend on the values
($\phi_{\infty}$) of moduli at spatial
infinity.\footnote{$\phi_{\infty}$ label different ground states (vacua) of the theory.}
Since the moduli are non-minimally coupled to gauge fields and the scalar charges are
non-zero at spatial infinity, one expects a modification of the first law of black hole
thermodynamics. That is, the first law of black hole thermodynamics should be supplemented
by a new term containing the variation of the moduli \cite{Gibbons:1996af}:
\begin{equation}
  dM=TdS+\psi^AdQ_A-\Sigma_i d\phi^i_{\infty},
\end{equation}
where, $\Sigma_i$ are the scalar charges and $\psi^A$ is the potential conjugate to the
$U(1)$ charges $Q_A$. Interestingly enough, the scalar charge is not protected by a gauge
symmetry and so it is not a conserved charge. Therefore, this form of the first law should
be taken with caution: in string theory the scalar fields (moduli) are interpreted as
local coupling constants and so a variation of the moduli values at infinity is equivalent
to changing the background.\footnote{As mentioned in \cite{Gibbons:1996af}, one could imagine, for
 example, a cosmological scenario in which $\phi_\infty$ does vary.}

Unlike the non-extremal case where the near horizon geometry (and the entropy) depends on
the values of the moduli at infinity, in the extremal case, the near horizon geometry is
universal and is determined by only the charge parameters. Consequently, the entropy is
also independent of the asymptotic values of the moduli. The scalars vary radially, but
they are `attracted' to fixed values at the horizon, depending only on the charge
parameters.\footnote{A similar behaviour was obtained for the rotating 
black holes \cite{Astefanesei:2006dd}. However, in this case the values of 
the scalars at the horizon have also an angular dependence.}

The attractor mechanism was discovered in the context of $4$-dimensional ${\cal N}=2$
supergravity \cite{Ferrara:1995ih, Strominger:1996kf, Ferrara:1996dd}, then extended to
other supergravities \cite{Ferrara:1996um} ---  including actions with higher derivative
corrections \cite{9812082}(see, e.g., \cite{Mohaupt:2000mj} for a nice review on this subject). It is
now well understood that supersymmetry does not really play a fundamental role in the
attractor phenomenon --- the attractor mechanism relies only on the form of the near
horizon geometry \cite{Sen:2005wa}. For spherically symmetric black holes, the near
horizon geometry is $AdS_2\times S^2$, which continues to ensure the attractor behaviour
even after including $\alpha'$ corrections
\cite{Sen:2005wa, Sen:2005iz, Sahoo:2006vz, Sahoo:2006rp, Alishahiha:2006ke,
  Chandrasekhar:2006kx}.
In fact, it has been suggested that, the `long throat' of $AdS_2$ is at the basis of
attractor mechanism \cite{Sen:2005wa, Astefanesei:2006dd, Kallosh:2006bt}.

The paper is organised as follows: in section \ref{sec:general}, we describe the general
set-up involving the action, equations of motion, and the effective potential. In section
\ref{sec:non-extremal}, we study the spherically symmetric non-extremal solutions,
focusing on some of the exact solutions. We explicitly discuss the appearance of scalar
charge in the context of first law of black hole thermodynamics. In section
\ref{attractorsection}, we discuss the attractor mechanism in the context of the extremal
limit of our black hole solutions.  We discuss the equivalence of the effective potential
approach \cite{Ferrara:1997tw,Goldstein:2005hq} and the entropy function \cite{Sen:2005wa}
formalism in the near horizon limit for a specific effective potential. We show that, for
the non-extremal black hole solutions whose near horizon geometry is not $AdS_2\times
S^2$,
the effective potential is not generically minimised and hence there is no attractor
behaviour.
We generalize the results of \cite{Goldstein:2005hq} to non-extremal black holes and discuss how the
relationship between the effective potential and entropy is modified.
We also discuss the role of attractor in the case of near-extremal black holes with large
charges in string theory. Using a perturbation analysis of non-extremal Reissner-Nordstrom
black holes with respect to variations of the asymptotic moduli, we explicitly find that,
to first order in perturbation theory, the near horizon geometry depends on the
moduli. Finally, we discuss the conditions for the attractor phenomenon in special
geometry language \cite{Ferrara:1997tw, Kallosh:2006bt} as well as the relation with the
other methods. In section \ref{sec:entropy.match} we discuss the role of attractor
mechanism in understanding the entropy of non-BPS extremal black holes.  For some examples
of non-supersymmetric extremal black holes in ${\cal N}=2$ four-dimensional supergravity
\cite{Kaplan:1996ev, Emparan:2006it} an agreement has been found between the
Bekenstein-Hawking entropy and the microscopic entropy computed in string theory.
It is tempting to conjecture that the deeper reason for this matching is the
  attractor mechanism. For $Ad_2\times S^2$ geometries, the equations of motion are
  equivalent to extremising the entropy function which fixes the moduli. The horizon is
  an attractor and so the near horizon geometry is universal, determined just by
  charges. One of the moduli is the dilaton that controls the Newton constant and so by
  moving from strong coupling to weak coupling asymptotically, the entropy determined by
  the near horizon geometry, does not change.\footnote{The
  authors of \cite{hep-th/0611143} also suggest that the attractor mechanism is at the
  root of this matching and carefully discuss numerous examples.}

\section{General set-up}\label{sec:general}

We will focus on a 4-dimensional theory of gravity coupled
to a set of massless scalars and vector fields, whose general bosonic action has the
form\footnote{We also consider `axionic' type couplings characterised by
  ${\tilde f}_{AB}$. In the so-called axion-dilaton-gravity model, the coupling is a
  $pseudo-scalar$ such that the action is parity-invariant. }
\begin{eqnarray}
  \nonumber
  I[G_{\mu\nu},\phi^i,A_{\mu}^I]
  \!&=&\!-
  \frac{1}{k^2}\int_{M} d^{4}x 
  \sqrt{-G}[ R-2g_{ij}(\phi)\partial_\mu\phi^i\partial^\mu\phi^j-f_{AB}(\phi)F^{A}_{\mu\nu}F^{B\, \mu\nu} \\
  &&-{\tilde f}_{AB}(\phi) F^A_{\mu \nu}
  F^B_{\rho \sigma} \epsilon^{\mu \nu \rho \sigma} ]  
  +\frac{2}{k^2}\int_{\partial M} d^{3} x \sqrt{-h}\Theta,
  \label{actiongen}
\end{eqnarray}
where $F^A_{\mu\nu}$, with $A=(0, \cdots N)$, are the $U(1)$ gauge fields, $\phi^i$, with
($i=1, \cdots, n$), are the scalar fields, $\epsilon^{\mu\nu\rho\sigma}$ is the completely
antisymmetric tensor, and $k^2=16\pi G_N$. The last term is the boundary Gibbons-Hawking
term; $h_{ab}$ and $\Theta$ are the induced metric and the trace of the extrinsic
curvature of the boundary geometry, respectively. The moduli determine the gauge coupling
constants and $g_{ij}(\phi)$ is the metric on the moduli space. We use Gaussian units so
that factors of $4\pi$ in the gauge fields can be avoided and the Newton's constant $G_N$ is
set to 1. The above action resembles that of the $ungauged$ supergravity
theories.

The equations of motion for the metric, moduli, and the gauge fields are given by

\begin{equation}
  R_{\mu\nu}-2g_{ij}\partial_{\mu}\phi^i\partial_{\nu}\phi^j 
  =f_{AB}\left(2F^A_{\phantom{A}\mu\lambda}F^{B\phantom{\nu}\lambda}_
    {\phantom{B}\nu}-
    {\textstyle \frac{1}{2}}G_{\mu\nu}F^A_{\phantom{a}\alpha\lambda} 
    F^{B \alpha\lambda} \right)\, , 
  \label{einstein}
\end{equation}
\begin{equation}
  \frac{1}{\sqrt{-G}}\partial_{\mu}(\sqrt{-G}g_{ij}\partial^{\mu}\phi^j)
  =\frac{1}{4} \frac{\partial f_{AB}}{\partial \phi^i} F^A _{\phantom{A}\mu\nu} 
  F^{B\, \mu\nu} 
  +\frac{1}{8}\frac{\partial \tilde f_{AB}}{\partial \phi^i} 
  F^A_{\phantom{A}\mu\nu} F^B_{\phantom{B}\rho \sigma} 
  \epsilon^{\mu\nu\rho\sigma} \, ,\\
  \label{dilaton}
\end{equation}
\begin{equation}
  \partial_{\mu}\left[\sqrt{-G}\left(f_{AB} F^{B\, \mu\nu} 
      + \frac{1}{2} {\tilde f}_{AB}F^B_{\phantom{B}\rho\sigma}
      \epsilon^{\mu\nu\rho\sigma}\right) \right] =  0\, ,
  \label{gaugefield}
\end{equation}
where we have varied the moduli and the gauge fields independently. The Bianchi identities
for the gauge fields are $F_{\phantom{A}\,[\mu\nu;\lambda]}^{A}=0$.

We consider the following  spherically symmetric  ansatz for the metric:
\begin{eqnarray}
  ds^{2} & = & -a(r)^{2}dt^{2}+a(r)^{-2}dr^{2}+b(r)^{2}d\Omega^{2}.
  \label{metric2}
\end{eqnarray}

The Bianchi identity and equation of motion for the gauge fields can be solved by a field
strength of the form \cite{Goldstein:2005hq}
\begin{equation}
  \label{fstrenghtgen}
  F^A=f^{AB}(Q_{B}-{\tilde f}_{BC}P^C) {1\over b^2} dt\wedge dr + 
  P^A \sin \theta  d\theta \wedge d\phi,
\end{equation}
where $P^A, Q_{A}$ are constants which determine the magnetic and electric charges carried
by the gauge field $F^A$ and $f^{AB}$ is the inverse of $f_{AB}$.

As discussed in \cite{Goldstein:2005hq}, given our ansatze the equations of motion can be
written as:
\begin{eqnarray}
  (a^{2} b^{2})^{''} & = & 2\, , \label{eq1} \\ 
  \frac{b^{''}}{b} & = & -g_{ij} \partial_{r}\phi^i 
  \partial_r\phi^j\, , \label{eq2}
\end{eqnarray}
\begin{equation}
  -1+a^{2}b^{'2}+\frac{a^{2'}b^{2'}}{2}=-\frac{1}{b^{2}}V_{eff}(\phi_i)+
  a^{2}b^{2}
  g_{ij} \partial_{r}\phi^i \partial_r\phi^j
  \, ,\label{constraint}
\end{equation}
\begin{equation}
  \label{eq3}
  \partial_{r}(a^{2}b^{2}g_{ij}\partial_{r}\phi^j)=\frac{1}{2b^{2}}
  \frac{\partial V_{eff}}{\partial \phi^i},
\end{equation}
where $'$ denotes derivatives with respect to $r$.
The first three equations come from the Einstein equations and the last one is the
equation of motion for the scalar. As discussed in Appendix
\ref{appendix1}, $V_{eff}(\phi^i)$ is a function of scalars fields, $\phi^i$, and charges,
$(Q_A,P^A)$ which  is given by
\begin{equation}
  \label{defpotgen}
  V_{eff}(\phi_i)=f^{AB}(Q_A-{\tilde f}_{AC}P^C)(Q_B - {\tilde f}_{BD}P^D)
  +f_{AB}P^AP^B.
\end{equation}
Modulo factors of $b^2$, one sees from (\ref{eq3}) that $V_{eff}(\phi^i)$ is an `effective
potential' for the scalar fields which is generated by non-trivial form fields. The
effective potential, first discussed in \cite{Ferrara:1997tw}, plays an important role in
describing the attractor mechanism
\cite{Goldstein:2005hq, Ferrara:2006yb, Kallosh:2006ib, Tripathy:2005qp, Denef:2000nb,
  Denef:2001xn, Kaura:2006mv}.

\section{Non-extremal black holes}\label{sec:non-extremal}

In this section, we study spherically symmetric non-extremal black holes for a model with
one scalar field non-minimally coupled to {\it two} gauge
fields.\footnote{A subset of these models, with $\alpha_1=-\alpha_2$, are equivalent to a
  system with {\it one} gauge field with both electric magnetic charges turned on.}
In certain cases, with exponential couplings, the equations of motion can be solved
exactly by rewriting them as Toda equations \cite{Gibbons:1987ps} (see Appendix
\ref{toda}).  Since we will consider exponential couplings, we need at least two terms in
the effective potential it to have a minimum which we required for the attractor
behaviour. Our solutions will be characterized by the mass, two gauge field charges,
values of the moduli at infinity and the scalar
charge.\footnote{The scalar charge is not really an independent parameter --- this is an
  example of secondary hair which is discussed further in sections \ref{scalar1law} and
  \ref{sec:discussion}.}

After discussing how to dress the charges and revisiting the equations of motion we will
discuss some constraints on the charges which follow from evaluating the equations at the
boundaries and then list some of the solutions.  Finally, we discuss, in detail, the role
of the scalar charge in the first law of thermodynamics for the simplest case from
subsection \ref{sec:exact}.

\subsection{Scaling symmetry and dressing the charges}
\label{sec:dressing}

We consider two gauge fields with modulus dependent  couplings of the form
\begin{equation}
  f_{AB}(\phi)=\delta_{AB}e^{\alpha_B \phi}
\end{equation}
for which exact solutions can be found. Given these couplings and taking the metric on the
modulus to one, the matter Lagrangian is
\begin{equation}
  \mathcal{L}_{matter}= 2 (\partial \phi)^2+ e^{\alpha_{1}\phi}(F_{1})^{2}+e^{\alpha_{2}\phi}
  (F_{2})^{2}.
  \label{l_gauge}
\end{equation}
For electrically charged solutions, assuming the ansatz (\ref{fstrenghtgen}), the
effective potential defined earlier becomes
\begin{equation}
  V_{eff}=e^{-\alpha_{1}\phi}(Q_{1})^{2}+e^{-\alpha_{2}\phi}(Q_{2})^{2}.
  \label{seffpot2}
\end{equation}
Alternatively for a dyonic black hole with a single gauge field we have
\begin{equation}
 V_{eff}=e^{-\alpha_{1}\phi}(Q_{1})^{2}+e^{\alpha_{1}\phi}(P_{1})^{2}.
\end{equation}  

The electric charges, $Q_A$, can be written in terms of the following surface integral at
spatial infinity:
\begin{equation}
\label{portocala}
  Q_A=\oint_{S^2_{\infty}}*f_{A B}F^B.
\end{equation}
From the equations of motion for the gauge fields, (\ref{gaugefield}), one observes that
this is the $U(1)$ charge one expects from Gauss' law modified in the presence of
moduli. On the other hand the Lagrangian (\ref{l_gauge}) is invariant under the global scaling
symmetry
\begin{equation}
  \label{eq:shift:sym}
  \phi' = \phi - \delta\phi\, , \qquad F_A' = e^{\alpha_A\delta\phi/2} F_A,
\end{equation}
but it is not hard to see that $Q_A$ is not invariant under this
symmetry. However,  one can define a dressed charge,
$\bar{Q}_A$,
\begin{equation}
  \label{eq:dressed:Q}
  \bar{Q}_A=e^{\frac{1}{2}\alpha_A \phi_\infty}\oint_{S^2_{\infty}}*f_{AB}F^B=Q_A\, 
  e^{\frac{1}{2}\alpha_A\phi_{\infty}},
\end{equation}
which is invariant --- The extra factor of $ e^{\frac{1}{2}\alpha_A\phi_\infty}$ absorbs
the change in $Q_A$.  Similarly, for magnetic charges, we can define the dressed charge
\footnote{A simple mnemonic for remembering how the dressed charges are defined is to check
  what factors are required to keep the effective potential invariant under rescaling.}
\begin{equation}
  \bar{P}_A=P_A\,e^{-\frac{1}{2}\alpha_A\phi_{\infty}}.
\end{equation}

\subsection{Equations of motion revisited}\label{sec:eom2}

Now, to recast the equations of motion as generalised Toda equations, and to facilitate some
of the thermodynamic analysis, we define the following new variables
\begin{equation}\label{new_vari}
  u_{1}=\phi\,, \qquad u_{2}=\log a\,, \qquad z=\log ab\,, \qquad 
  \mbox{``}\cdot\mbox{''}=\partial_{\tau}=a^{2}b^{2}\partial_{r}.
\end{equation}
From the definition of $\tau$, we find
\begin{eqnarray}
  \label{eq:rtau}
  r&=& \frac{r_+e^{-c\tau}-r_-e^{c\tau}}{e^{-c\tau}-e^{c\tau}}\,,\quad 
  \mbox{where}
  \quad 
  c=\half(r_+-r_-)
\end{eqnarray}
and conversely
\begin{eqnarray}
  \label{eq:tau}
  \tau&=&\frac{1}{(r_{+}-r_{-})}\log\left(\frac{r-r_{+}}{r-r_{-}}\right)\,.
\end{eqnarray}
In terms of these new variables, the equations of motion become
\begin{eqnarray}
  \ddot{u}_{1} & = & \frac{1}{2}\alpha_{1}e^{2u_{2}+
    \alpha_{1}u_{1}}Q_{1}^{2}+\frac{1}{2}\alpha_{2}e^{2u_{2}+
    \alpha_{2}u_{1}}Q_{2}^{2}\, ,
  \label{eq:phi_eom_3}\\
  \ddot{u}_{2} & = & e^{2u_{2}+\alpha_{1}u_{1}}Q_{1}^{2}+e^{2u_{2}+
    \alpha_{2}u_{1}}Q_{2}^{2}\, ,
  \label{eq:v_eom_1}\\
  \ddot{z} & = & e^{2z}\, ,
  \label{eq:z_EOM}\\
  \dot{z}^{2}-e^{2z} & = & \dot{u_{1}}^{2}+\dot{u_{2}}^{2}-e^{2u_{2}+
    \alpha_{1}u_{1}}Q_{1}^{2}-e^{2u_{2}+\alpha_{2}u_{1}}Q_{2}^{2}.
  \label{eq:toda_E}
\end{eqnarray}
The last equation is equivalent to the Hamiltonian constraint, (\ref{constraint}).  The
equation for $z$, (\ref{eq:z_EOM}), decouples from the other equations and
is equivalent to (\ref{eq1}).  It can be used to show that the left hand side of the
Hamiltonian constraint, (\ref{eq:toda_E}), is a constant, since
\begin{equation}
\partial_\tau(\dot z^2-e^{2z})=2\dot z (\ddot z - e^{2z})\refeq{eq:z_EOM}0.   
\end{equation}
So, letting $c^2=\dot z^2 - e^{2z}$, we can rewrite (\ref{eq:toda_E}) as
\begin{equation}
  c^2  =  \dot{u_{1}}^{2}+\dot{u_{2}}^{2}-e^{2u_{2}+
    \alpha_{1}u_{1}}Q_{1}^{2}-e^{2u_{2}+\alpha_{2}u_{1}}Q_{2}^{2}.
  \label{eq:ham2}
\end{equation}
The constant $c$ above, turns out to be the same as the one we defined in (\ref{eq:rtau}). 
 
\subsection{Constraints on the charges}

Examining the equations of motion evaluated at the boundaries one finds
two important properties of these black holes:
\begin{eqnarray}
  M^2+g_{ij}(\phi_\infty)\Sigma^i\Sigma^j-V_{eff}(\phi_\infty)=4S^2T^2=c^2\, ,
  \label{general:niceconstraint}\\
  2g_{ij}(\phi_\infty)M\Sigma^j-2g_{ij}(\phi_\infty){K^j}-g_{ij,k}(\phi_\infty)\Sigma^j\Sigma^k= 
  -\frac{1}{2}\frac{\partial V_{eff}}{\partial \phi^i_\infty}\, , 
  \label{general:sigma}
\end{eqnarray}
where, $S$ and $T$ are the entropy and temperature of the black hole, and the scalar
monopole and dipole charges, $\Sigma$ and $K$, are defined by the expansion of the moduli
at infinity:
\begin{equation}
\phi^j = \phi^j_{\infty} + \frac{\Sigma^j}{r} + \frac{K^j}{r^2}+\ldots\;.
\end{equation}
The first equation, (\ref{general:niceconstraint}), is the Hamiltonian constraint
(\ref{eq:ham2}) evaluated at the boundaries and it provides a constraint on charges. The
second one, (\ref{general:sigma}), is an expression of the dependence of the scalar charges
on the mass, the electric and magnetic charges and the values of the moduli at infinity. It
follows from evaluating the scalar field equation at infinity.

For simplicity, we prove the relations, (\ref{general:niceconstraint}) and
(\ref{eq:sigma:proof}), in the case of two charges and one scalar field, but the argument
is easy to generalise. To prove (\ref{general:niceconstraint}), we evaluate the right hand
side of (\ref{eq:ham2}) at spatial infinity and evaluate the constant, $c^2$, at the
horizon.  We find that
\begin{equation}
  c^2=\dot{z}^{2}-e^{2z}=\left[ab\partial_{r }(ab) \right]^2-e^{2\ln(ab)}=
  (a a' b^2)^2=4S^2T^2\, .
\end{equation}
Here we used the fact that, for a non-extremal black hole with finite horizon area, $a(r)$
has a simple zero and $b$ is a constant at the horizon. In addition, we use the following
expressions for the temperature and entropy:
\begin{equation}
  T=\left.\frac{a a'}{2\pi}\right|_{r=r_h}\neq 0  
  \,,\quad S=\pi b^2(r_h).
\end{equation}
Then, evaluating the right hand side of (\ref{eq:ham2}) at infinity gives the left hand side of
(\ref{general:niceconstraint}).  To prove (\ref{general:sigma}), we consider the equation
of motion for the scalar field, (\ref{eq3}), evaluated at infinity. Upon simplification,
one finds, to first non-trivial order in $1/r$:
\begin{equation}
  \left[
    g_{ij}\left(1-\frac{2M}{r}\right) r^2 \left(\phi^j_\infty+ 
      \frac{\Sigma^j}{r}+
      \frac{K^j}{r^2}\right)'
  \right]'
  = \left.\frac{1}{2r^2}\frac{\partial V_{eff}}{\partial \phi^i_\infty}\right.
  \label{eq:sigma:proof}
\end{equation}
which  leads to (\ref{general:sigma}).

\subsection{Exact solutions}
\label{sec:exact}

It is only for certain values of the parameters, $\alpha_A$, that exact
solutions are known. The parameter, $\gamma$, given by
\begin{equation}
  \gamma=\tfrac{1}{2}\left(\sqrt{1-2\alpha_{1}\alpha_{2}}-1\right)\, 
\end{equation} 
is useful for characterising the solutions: exact solutions are know for the cases
$\gamma=1,2,3$ \cite{Gibbons:1987ps,Goldstein:2005hq}.  The details of constructing some exact non-extremal solutions are
discussed in Appendix \ref{fsolutions}. We merely list some of the solutions below for the
cases $\gamma=1,2$.

\subsubsection{Case Ia: $\gamma=1$ and $|\alpha_{1,2}|=2$}
\label{cazul}

This solution has been extensively discussed in the literature. Since, in subsection
\ref{scalar1law}, we wish to discuss the role of both electric and magnetic charges in the
first law, we present this solution as a dyonic black hole with a single gauge field so, 
\begin{equation}
  V_{eff}=e^{-2\phi}Q^2+e^{2\phi}P^2.
\end{equation}
The non-extremal solution is given by \cite{Kallosh:1992ii}:
\begin{eqnarray}
  \exp(2\phi) & = & e^{2\phi_{\infty}}\frac{(r+\Sigma)}{(r-\Sigma)}\, ,
  \nonumber \\
  a^{2} & = & \frac{(r-r_{+})(r-r_{-})}{(r^{2}-\Sigma^{2})}\, ,
  \label{eq:alpha2_solution}\\
  b^{2} & = & (r^{2}-\Sigma^{2}),\nonumber 
\end{eqnarray} 
with
\begin{equation} r_{\pm}=M\pm c\,, \qquad
  c=\sqrt{M^{2}+\Sigma^{2}-\bar{Q}^{2}-\bar{P}^{2}},\label{rhne}
\end{equation}
where we have defined the dressed charges $\bar Q= e^{\phi_\infty}Q$ and  $\bar P= e^{-\phi_\infty}P$.

As already discussed, the scalar charge, $\Sigma$, is not an independent
parameter. It is given by
\begin{equation}
  \Sigma=\frac{\bar{P}^{2}-\bar{Q}^{2}}{2M}.
\end{equation}
The extremal limit of the above solution corresponds to letting $c^2\rightarrow0$ and the
corresponding solution can be embedded in ${\cal N}=4$ supergravity.

\subsubsection{Case Ib: $\gamma=1$ , $\sqrt{-\alpha_1\alpha_2}=2$ and $\alpha_1>\alpha_2$
}\label{sec:case-ib}

For this case,
\begin{equation}
  V_{eff}=e^{-\alpha_1\phi}Q_1^2+e^{4\phi/\alpha_1}Q_2^2.
\end{equation}
The solution is
\begin{equation}
  \begin{array}{cll}
    e^{(\alpha_{1}-\alpha_{2})\phi} 
    & = &  
    \left(\frac{1-\lambda}{\lambda}\right)\left(\frac{Q_{2}F_{2}}
      {Q_{1}F_{1}}\right)^{2}\\
    a^{2} & = &  
    \left.
      c^2\left(Q_{1}F_{1}\right)^{-2(1-\lambda)}
      \left(Q_{2}F_{2}\right)^{-2\lambda}
    \right/
    \left\{ \left(\frac{1-\lambda}{\lambda}\right)^{\lambda}+
      \left(\frac{1-\lambda}{\lambda}\right)^{1-\lambda}\right\}\\
    b^2 &=&(r-r_+)(r-r_-)/a^2
  \end{array}
  \label{eq:exact_sol_case1}
\end{equation}
where
\begin{equation}
  \label{eq:def:Fi}
  F_i  =  \sinh(c(\tau-d_i))
\end{equation}
and the ratio $\lambda$ is defined as
\begin{equation}
  \label{eq:def:lambda}
  \lambda  =  \frac{\alpha_{1}}{\alpha_{1}-\alpha_{2}}.
\end{equation}
Notice that $ \lambda$ lies between $0$ and $1$. The integration constants $c$, $r_\pm$,
and $d_i$ are given in terms of $M$, $\Sigma$, and the `dressed' charges 
\begin{equation}
  \label{eq:q1bar:def}
  \bar{Q_i}^2=e^{\alpha_i\phi_\infty}Q_i^2 \, 
\end{equation}
by
\begin{equation}
  \begin{array}{clll}
    c^2 &=& M^2 +\Sigma^2 
    -{\bar Q}_1^2
    -{\bar Q}_2^2\, ,\\
    r_\pm&=&M\pm c\, ,
  \end{array}
\end{equation}
\begin{eqnarray}
  \label{eq:q1q2}
  \sinh^2(cd_{i})  =  
  \left[\frac{4 {c^2}}{\alpha_i^2+4} \right]\bar Q_{i}^{-2}\, .
\end{eqnarray}

Due to the fact that the parameters $\alpha_1$ and $\alpha_2$ are very weakly 
constrained, we find it unlikely that all solutions in this class could be embedded in 
supergravity theory. 

\subsubsection{Case II: $\gamma=2$ and $|\alpha_{1,2}|=2\sqrt{3}$}\label{sec:case-ii}
This case arises from the Kaluza-Klein reduction of the $5d$ Schwarzschild black hole so
it is natural to write it as a dyonic  solution with
\begin{equation}
   V_{eff}=e^{-2\phi/\sqrt3}Q^2+e^{2\phi/\sqrt3}P^2.
\end{equation}
  The solution can
be written as \cite{Dobiasch:1981vh,Gibbons:1985ac}
\begin{eqnarray}
  \exp(4\phi/\sqrt{3}) & =e^{4\phi_{\infty}/\sqrt{3}} & \frac{A}{B}\\
  a^{2} & = & \frac{(r-r_{+})(r-r_{-})}{\sqrt{AB}}\\
  b^{2} & = & \sqrt{AB}
\end{eqnarray}
\begin{eqnarray}
  A & = & (r-r_{A_{+}})(r-r_{A_{-}})\\
  B & = & (r-r_{B_{+}})(r-r_{B_{-}})
\end{eqnarray}
\begin{equation}
  r_{\pm}=M\pm\sqrt{M^{2}+\Sigma^{2}-\bar{P}^{2}-\bar{Q}^{2}}=M\pm c
\end{equation}
\begin{equation}
  r_{A_{\pm}}=\frac{1}{\sqrt{3}}\Sigma\pm\bar{P}\sqrt{\frac{2\Sigma}
    {\Sigma-\sqrt{3}M}}
\end{equation}
\begin{equation}
  r_{B_{\pm}}=-\frac{1}{\sqrt{3}}\Sigma\pm\bar{Q}\sqrt{\frac{2\Sigma}
    {\Sigma+\sqrt{3}M}}
\end{equation}
\begin{equation}
  \mathrm{Area}=4\pi\sqrt{(r_{+}-r_{A_{+}})(r_{+}-r_{A_{-}})
    (r_{+}-r_{B_{+}})(r_{+}-r_{B_{-}})}
\end{equation}
where we have defined the dressed charges $\bar Q= e^{\frac{1}{\sqrt3}\phi_\infty}Q$ and
$\bar P= e^{-\frac{1}{\sqrt3}\phi_\infty}P$.

Once again $\Sigma$ is not an independent parameter and is given by,
\begin{equation}
  \frac{2}{\sqrt{3}}\Sigma=\frac{\bar{Q}^{2}}{\sqrt{3}M+\Sigma}-
  \frac{\bar{P}^{2}}{\sqrt{3}M-\Sigma}.
\end{equation}

In the extremal limit ($c=0$) we obtain a non-BPS black hole that can be embedded in ${\cal N}=2$
supergravity.

\subsection{Scalar charge and the first law of thermodynamics}
\label{scalar1law}
We have seen previously that, unlike in the case of minimally-coupled
scalars\footnote{For minimally-coupled scalars the standard no-hair theorems apply and do
  not allow for a nice solution with non-zero scalar charge. },
the black hole solutions we consider carry {\it scalar charge}.  It is important to
mention that the scalar charge is not protected by a gauge symmetry, and hence is not a
conserved charge.

In the cases we studied, the scalar charge is not an independent parameter.  It depends on
the other asymptotic charges, namely the ADM mass and the dressed gauge field
charges. This implies that just one of the parameters $\phi_{\infty}$ and $\Sigma$ is
independent. This kind of scalar charge, which depends on other asymptotic data , is
called {\it secondary hair}. As it depends on the gauge field charges, it does not
represent a new quantum number associated with the black hole.

Due to the non-minimal coupling of the scalar fields, the first law gets modified. It
should be supplemented by a new term containing the variation of the moduli
\cite{Gibbons:1996af}:
\begin{equation}
  \label{1law}
  dM=TdS+\psi^AdQ_A+\psi_AdP^A-\Sigma_i d\phi^i_{\infty}\, ,
\end{equation}
where $(\psi^A ,\psi_A)$ are the potentials conjugate to the charges $(Q_A ,P^A)$.

Indeed, we will show explicitly that this is the case for some of the black hole solutions
considered here. We need to check that $M(S,Q_A,P^A,\phi_{\infty})$ is an exact
differential.  Since it is particular to this class of black holes, we are mainly
interested in the non-trivial term
$\Sigma_i=-(\partial M/\partial \phi^i_{\infty})|_{(S,Q_A,P^A)}$, but similar computations
can be done for the other terms.
\footnote{Note that, requiring cosmic censorship imposes the condition $M\geq \Sigma$
  since, there is a curvature singularity at $r=\Sigma$ while the outer horizon is at
  $r_+=M+c$ (which approaches $M$ as $c\rightarrow0$).}

While we have verified the first law for all the solutions in the previous
section, we present, in detail, the analysis of the solution (Ia), which has $\gamma=1$ and
$\alpha_{1}=-\alpha_{2}=2$, from section \ref{cazul}. In that case, there are two
conserved gauge charges $(P, Q)$ and a scalar charge. We rewrite some of the equations
from subsection \ref{cazul}, as well as the entropy, in the following useful way:
\begin{eqnarray}
  2M\Sigma&=&\bar{P}^2-\bar{Q}^2\, ,
  \label{good}
  \\
  c&=&\sqrt{\left(M+\frac{\bar{P}^{2}-\bar{Q}^{2}}{2M}\right)^2-2\bar{P}^{2}}\, ,
  \label{goodconstantN}
  \\
  S&=&\pi b^2(r_+)=\pi[(c+M)^2-\Sigma^2]=\pi(2M^2-\bar{Q}^{2}-\bar{P}^{2}+2cM)\, .
  \label{goodentropyN}
\end{eqnarray}

Now, by differentiating (\ref{goodconstantN}) and (\ref{goodentropyN}) at fixed entropy
and charge parameters $Q, P$, we obtain:
\begin{eqnarray}
  dc&=&\frac{1}{c}(M+\Sigma)\left( 1-\frac{\Sigma}{M} \right) dM + 
  \frac{1}{c}\left[ 2\bar{P}^2-
    \frac{M+\Sigma}{M}\left( \bar{P}^2+\bar{Q}^2 \right) 
  \right]d\phi_{\infty}\, , 
  \label{constdif}
  \\
  0&=&4MdM-2\bar{Q}^{2}d\phi_{\infty}+2\bar{P}^{2}d\phi_{\infty}+2cdM+2Mdc\, .
  \label{entropydif}
\end{eqnarray}
The next step is to use (\ref{constdif}) in (\ref{entropydif}) and to check that, indeed,
$\Sigma=-(\partial M/\partial \phi_{\infty})|_{S,Q,P}$.  We can calculate the other
intensive parameters in the same way and we obtain the following expressions:
\begin{equation}
  \psi^Q=\frac{\bar{Q}^2}{Q}\frac{M+\Sigma+c}{(M+c)^2-\Sigma^2}\, ,  \qquad
  \psi^P=\frac{\bar{P}^2}{P}\frac{M+c-\Sigma}{(M+c)^2-\Sigma^2}\, .
\end{equation}
With all these expressions one can easily check that the first law is satisfied:
\begin{equation}
  \label{1lawnaspa}
  dM=TdS+\psi^QdQ+\psi^PdP-\Sigma_i d\phi^i_{\infty}.
\end{equation}

The fact that the Lagrangian has a global scaling symmetry, (\ref{eq:shift:sym}), suggests
writing the first law in terms of the `dressed' charges:
\begin{equation}
  \label{good1law}
  dM=TdS+\bar{\psi}^Qd\bar{Q}+\bar{\psi}^Pd\bar{P},
\end{equation}
where $\bar{\psi}^Q$ and $\bar{\psi}^P$ are the conjugate potentials of the dressed
charges $\bar{Q}$ and $\bar{P}$, respectively.  One can again compute the values of the
intensive parameters as above or one can rewrite (\ref{good1law}) as
\begin{equation}
  dM=TdS+\bar{\psi}^Q\frac{\bar{Q}}{Q}dQ+
  \bar{\psi}^P\frac{\bar{P}}{P}dP+(\bar{\psi}^Q\bar{Q}-\bar{\psi}^P\bar{P})
  d\phi_{\infty},
\end{equation} 
and then compare with (\ref{1lawnaspa}). One obtains the following expressions for the
intensive parameters:
\begin{equation}
  \bar{\psi}^Q=\frac{\bar{Q}}{M+c-\Sigma}, \qquad
  \bar{\psi}^P=\frac{\bar{P}}{M+c+\Sigma}.
\end{equation}

As an application of our formulae consider what happens if we add some scalar particles to
the black hole. We keep $\bar Q$ and $\bar P$ (or equivalently $Q$, $P$ and $\phi_\infty$)
fixed. Taking a differential of (\ref{good}) and using the first law with $\bar Q$ and
$\bar P$ constant gives
\begin{equation}
  \label{eq:scalar_abs}
  \delta S=-\frac{M}{T}\frac{\delta\Sigma}{\Sigma}.
\end{equation}
From (\ref{eq:scalar_abs}) we see that increasing $|\Sigma|$ (i.e. when $d\Sigma/\Sigma>0$)
causes the entropy to decrease. This implies that adding scalar charge induces Hawking
radiation. Conversely, reducing $|\Sigma|$ simply causes the black hole to puff up.

Note that (\ref{good1law}) does not involve a variation with respect to the asymptotic
value of the scalars. However, let us recall that the physical conserved charges (due 
to the equations of motion) are $Q_A$ given in (\ref{portocala}) and so the scaling symmetry 
does not preserve the conserved charges. By making a scaling one can generate new solutions. 
However, the new solution can not be reached {\it dynamically} starting from the old one because this 
will also force a violation of charge conservation. To obtain the first law in this form, one should 
supplement the quasilocal formalism by a boundary counterterm that depends of moduli --- this term 
is taking care of the non-conserved charge in the first law. 

\section{Attractor mechanism}\label{attractorsection}

In this section we discuss the attractor mechanism using both, the effective potential
(\ref{defpotgen}) method \cite{Goldstein:2005hq} and the entropy function
\cite{Sen:2005wa} framework. The first method is based on investigating the equations of
motion of the moduli and finding the conditions satisfied by the effective potential such
that the attractor phenomenon occurs. We will extend the calculations of
\cite{Goldstein:2005hq} for non-extremal black holes. The entropy function
approach focuses on the near-horizon geometry and its enhanced symmetries.  The
equivalence of the effective potential approach and entropy function formalism in the
context of four-dimensional extremal non-BPS black hole solutions in ${\cal N} = 2$
supergravity has recently been discussed in \cite{Cardoso:2006cb}.  In the last subsection
we shall briefly mention the relevant aspects of the attractor phenomenon in special
geometry \cite{Ferrara:1997tw, Kallosh:2006bt}.

\subsection{Effective potential and non-supersymmetric attractors}
\label{veff}
We consider again the solution (Ia) and we investigate its extremal limit, $c\rightarrow0$. We
use (\ref{good})-(\ref{goodentropyN}) and in the extremal limit we obtain:
\begin{eqnarray}
  2M\Sigma&=&\bar{P}^2-\bar{Q}^2\, ,
  \label{goodextremal}
  \\
  0&=&\sqrt{\left(M+\frac{\bar{P}^{2}-\bar{Q}^{2}}{2M}\right)^2-2\bar{P}^{2}}=
  \sqrt{\left(M+\Sigma \right)^2-2\bar{P}^{2}}\, ,
  \label{goodconstant}
  \\
  S&=&\pi b^2(r_+)=\pi \left(M^2-\Sigma^2\right)\, .
  \label{goodentropyextremal}
\end{eqnarray}
One can easily solve the system of the first two equations to obtain $M$ and $\Sigma$ as
functions of the dressed charges $\bar{Q}=e^{\phi_{\infty}}Q$ and
$\bar{P}=e^{-\phi_{\infty}}P$. Then, the mass and the scalar charge  depend on  the
asymptotic values of the moduli $\phi_{\infty}$ as follows,
\begin{eqnarray}
  M &=& \frac{1}{\sqrt 2} (\bar P + \bar Q) \, ,
  \label{mass}
  \\
  \Sigma &=&  \frac{1}{\sqrt 2} (\bar P - \bar Q) \, .
  \label{charge}
\end{eqnarray}
However, the entropy becomes independent of $\phi_{\infty}$, {\it i.e.}
\begin{eqnarray}
  S &=& 2\pi P Q\, . 
  \label{indentropy}
\end{eqnarray} 
The entropy is also independent of moduli for the other solutions.  In what follows, we
briefly review the effective potential method of \cite{Goldstein:2005hq} to clarify these
interesting results.

For the attractor phenomenon to occur, it is sufficient if the following two conditions
are satisfied \cite{Goldstein:2005hq}. Firstly, for fixed charges, as a function of the
moduli, $V_{eff}$ must have a critical point. Denoting the critical values for the scalars
as $\phi^i=\phi^i_0$ we have,
\begin{equation}
  \label{critical}
  \partial_iV_{eff}(\phi^i_{0})=0.
\end{equation}
Secondly, there should be no unstable directions about this minimum, so the matrix of second
derivatives of the potential at the critical point,
\begin{equation}
  \label{massmatrix}
  M_{ij}={1\over 2} \partial_i\partial_jV_{eff}(\phi^k_{0})
\end{equation}
should have no negative eigenvalues.  Schematically we can write,
\begin{equation}
  \label{positive}
  M_{ij}>0.
\end{equation}
The eigenvalues of $M_{ij}$ are proportional to the effective mass squared for the fields,
$\phi^i$, near the attractor point.

We can consistently set the moduli to constants if we fix them at their critical
values. The theory effectively reduces to Einstein-Maxwell gravity which has the extremal
Reissner-Nordstrom black hole as a solution. If we then examine what happens if the
asymptotic value of the asymptotic moduli deviate slightly from the attractor value,
simultaneously demanding that the black hole remains extremal (that is the horizon still
has a double zero), one finds attractor behaviour: the moduli attain their critical values
at the horizon and entropy remains independent of the value of the moduli at infinity
\cite{Goldstein:2005hq}. The horizon radius is given by
\begin{equation}
  \label{RH}
  b_H^2=V_{eff}(\phi^i_{0}),
\end{equation}
and the entropy is
\begin{equation}
  \label{BH}
  S_{BH}={1 \over 4} A = \pi b_H^2=\pi V_{eff}(\phi^i_{0}). 
\end{equation}

Now, one can verify the attractor behaviour for our effective potential given in
(\ref{seffpot2}).  The condition for the existence of an extremum for the effective
potential will give us the value of the moduli at the horizon :
\begin{equation}
  0=\frac{\partial V_{eff}}{\partial \phi}=
  \alpha_{1}e^{\alpha_{1}\phi}(Q_{1})^{2}+\alpha_{2}e^{\alpha_{2}\phi}
  (Q_{2})^{2} \Longrightarrow 
  e^{(\alpha_{1}-\alpha_{2})\phi_0}=-\frac{\alpha_{2}}{\alpha_{1}}
  \frac{Q_2^2}{Q_1^2}.
\end{equation}
Then, the value of the moduli at the horizon depends just on the charge parameters and not
on the boundary conditions. It is a simple exercise to check that the second derivative of
the potential is positive and hence the extremum is a minimum. For
$\alpha_{1}\neq -\alpha_{2}$ with $\alpha_1\alpha_2=-4$ the entropy is
\begin{equation}
  S=\pi V_{eff}(\phi_0)=\pi Q_1^{\frac{8}{\alpha_1^2+4}}
  Q_2^{\frac{2\alpha_1^2}{\alpha_1^2+4}}
  \left[\left(\frac{4}{\alpha_1^2}\right)^{\frac{\alpha_1^2}{\alpha_1^2+4}} + 
    \left(\frac{4}{\alpha_1^2}\right)^{-\frac{4}{\alpha_1^2+4}}\right]\, .
\end{equation}
The same result is obtained by taking the extremal limit $c=0$ in the non-extremal entropy
(\ref{misto1}).  For $\alpha_{1}=-\alpha_{2}=2$, the entropy is given by
\begin{equation}
  S=\pi V_{eff}(\phi_0)=2\pi Q_1Q_2\,,
\end{equation}
or for a dyonic black hole
\begin{equation}
  S=2\pi PQ\,.
\end{equation}

It is important to note that in deriving the conditions for the attractor phenomenon, one
does not have to use supersymmetry at all.  We will obtain the same result in the next
subsection by using the entropy function.

\subsection{Entropy function}
\label{efunction}
 
The near-horizon geometry of the extremal charged black holes has been shown to have a
geometry of $AdS_2\times S^2$ and, when embedded in certain supergravities, has an
enhanced supersymmetry.

As has been discussed in \cite{Ferrara:1997yr, Kallosh:2006bt}, the moduli do not preserve
any memory of the initial conditions at infinity due to the presence of the infinite
throat of $AdS_2$. This is in analogy with the properties of the 
behavior of dynamical flows in dissipative systems, where, on 
approaching the attractors, the orbits 
practically lose all the memory of their initial conditions.

Let us investigate the near-horizon geometry of non-extremal 
spherically symmetric black holes, where the line element 
is given by,
\begin{eqnarray}
  ds^{2} & = & -a(r)^{2}dt^{2}+a(r)^{-2}dr^{2}+b(r)^{2}d\Omega^{2}.
\end{eqnarray}
The Einstein equation (\ref{eq1}), $(a^2b^2)^{''}=2$, can be integrated out and one gets
$a^2b^2=(r-r_+)(r-r_-)$.  The interpretation of the parameters $r_+$ and $r_-$ is that
they are related to the outer and the inner horizon, respectively. Next, we introduce the
non-extremality parameter $\epsilon$ and also make a change of coordinates such that the
horizon is at $\rho=0$, {\it i.e.}
\begin{equation}
  \rho=r-r_+\,, \qquad \epsilon=r_+-r_-\,.
\end{equation}
The extremal black hole is obtained when the inner and the outer horizons coincide. For
the non-extremal solution ($r_+\neq r_-$), we have,
\begin{equation}
  a^2b^2=\rho(\rho+\epsilon)\, ,
\end{equation}
Let us take,
\begin{equation}
  \label{a}
  a^2=\rho f(r)=\rho(f_0+f_1\rho+f_2\rho^2+...)\, ,
\end{equation}
\begin{equation}
  \label{b}
  b^2=\frac{\rho(\rho+\epsilon)}{a^2}=\frac{\rho+\epsilon}{f_0+f_1\rho+
    f_2\rho^2+...}\, ,
\end{equation}
where, $f(r)$ has been expanded as a power series in $\rho$. The near-horizon geometry is
obtained by taking the limit $\rho\rightarrow 0$ and is given by
\begin{equation}
  \label{nonextreme}
  ds^2=-(\rho f_0)dt^2+\frac{1}{\rho f_0}d\rho^2+
  \frac{\epsilon}{f_0}d\Omega^2.
\end{equation}
The temperature and the entropy of the non-extremal black hole are given by
\begin{equation}
  \left.T=\frac{(a^2)^{'}}{4\pi}\right|_{\rho=0}, \qquad 
  \left.S=\frac{4\pi b^2}{4}\right|_{\rho=0}.
\end{equation}
By comparing these expressions with the expressions obtained from the near-horizon
geometry (\ref{nonextreme}), one can read off the following expressions for the parameters
appearing in (\ref{nonextreme}):
\begin{equation}
  f_0=4\pi T, \qquad \epsilon=\frac{f_0S}{\pi}=4TS=2c .
\end{equation}
To see that the near-horizon geometry of the extremal solution is $AdS_2\times S^2$, we
take the extremal limit $T\sim f_0 \rightarrow 0$ and expand the metric to first
non-trivial order about $\rho=0$. It is not hard to see that this procedure gives:
\begin{equation}
  ds^2=\frac{1}{f_1}(-\rho^2dt^2+\frac{1}{\rho^2}d\rho^2)+
  \frac{1}{f_1}d\Omega^2,
\end{equation}
where we have rescaled the time variable: $t\rightarrow t/f_1$.  

It is crucial that the near horizon geometry is $AdS_2\times S^2$: Sen's entropy function
formalism, \cite{Sen:2005wa}, assumes from the beginning that the metric and all other
fields respect the $SO(2,1)\times SO(3)$ symmetry of $AdS_2\times
S^2$.\footnote{In
  the rotating case, due to the axial symmetry, the $SO(3)$ symmetry is broken to a $U(1)$
  symmetry. However, the long throat of $AdS_2$ is still present and the entropy function
  formalism, slightly modified, can still be applied \cite{Astefanesei:2006dd}.}

In \cite{Sen:2005wa}, it was observed that the entropy of a spherically symmetric extremal
black hole is given by the extremum of the Legendre transform (with respect to the
electric field) of the Lagrangian density evaluated at the horizon.  The derivation of
this result does not require that the theory or the solution are supersymmetric.  The only
requirements are gauge and general coordinate invariance of the action and the assumption
that the near horizon geometry is $AdS_2\times S^2$.

The entropy function is defined as
\begin{equation}
  \F(\overrightarrow{u},\overrightarrow{v},\overrightarrow{e},
  \overrightarrow{p})
  =
  2\pi\left(e_iq_i-f(\overrightarrow{u},\overrightarrow{v},
  \overrightarrow{e},\overrightarrow{p})\right)
  =
  2\pi\left(e_iq_i-\int_H d\theta d\phi\sqrt{-G}{\cal L}\right),
\end{equation}
where $q_i=\partial f/\partial e_i$ are the electric charges, $u_s$ are the values of the
moduli at the horizon, ${p_i}$ and ${e_i}$ are the near horizon radial magnetic and
electric fields and $v_1$, $v_2$ are the sizes of $AdS_2$ and $S^2$ respectively. Thus,
$\F/2\pi$ is the Legendre transform of the reduced Lagrangian, $f$, with respect to the
variables $e_i$. For an extremal black hole of electric charge $\overrightarrow{Q}$ and
magnetic charge $\overrightarrow{P}$, Sen has shown that the equations determining
$\overrightarrow{u},\overrightarrow{v}$, and $\overrightarrow{e}$ are given by:
\begin{equation}
  \frac{\partial \F}{\partial u_s}=0\,, \qquad 
  \frac{\partial \F}{\partial v_i}=0\,, 
  \qquad \frac{\partial \F}{\partial e_i}=0\,.
  \label{attractor}
\end{equation}
Then, the black hole entropy is given by
$S=\F(\overrightarrow{u},\overrightarrow{v},\overrightarrow{e}, \overrightarrow{p})$ at the
extremum (\ref{attractor}). The entropy function,
$\F(\overrightarrow{u},\overrightarrow{v},\overrightarrow{e}, \overrightarrow{p})$,
determines the sizes $v_1$, $v_2$ of $AdS_2$ and $S_2$ and also the near horizon values of
moduli ${u_s}$ and gauge field strengths ${e_i}$. If $\F$ has no flat directions, then the
extremization of $\F$ determines $\overrightarrow{u}$, $\overrightarrow{v}$,
$\overrightarrow{e}$ in terms of $\overrightarrow{Q}$ and $\overrightarrow{P}$. Therefore,
$S=\F$ is independent of the asymptotic values of the scalar fields.  These results lead to
a generalised attractor phenomenon for both supersymmetric and non-supersymmetic extremal
black hole solutions.

Now we can apply this method to our action (\ref{actiongen}) with a zero axionic
coupling. We are interested in a theory with one scalar field and one electromagnetic
field with both electric and magnetic charges turned on:
\begin{equation}
  S=\frac{1}{\kappa^{2}}\int d^{4}x\sqrt{-G}(R-2(\partial\phi)^{2}-
  e^{2\phi}F^{2}).
\end{equation}
The general metric of $AdS_2\times S^2$ can be written as
\begin{equation}
  ds^2=v_1\left(-\rho^2dt^2+\frac{1}{\rho^2}d\rho^2\right)+
  v_2(d\theta^2+\sin^2\theta d\phi^2)\, .
\end{equation}
The field strength ansatz (\ref{fstrenghtgen}) in our case is given by
\begin{equation}
  F=edt\wedge dr+P\sin\theta d\theta \wedge d\phi=e^{-2\phi}Qdt\wedge dr+
  P\sin\theta d\theta \wedge d\phi.
\end{equation}
The entropy function, $\F(v_1, v_2, e, q, p)$, and reduced Lagrangian, $f(v_1, v_2, e,
p)$, are given by
\begin{eqnarray}
  && \F(v_1,v_2,e,q,p)=2\pi [qe-f(v_1,v_2,e,p)]\, ,\\
  \nonumber
  && f(v_1,v_2,e,p)=\frac{8\pi}{k^2}\left[-v_2+v_1-
    e^{2\phi}\left(\frac{-v_2}{v_1}e^2+\frac{v_1}{v_2}P^2\right)\, \right]\, .
\end{eqnarray}
Then the attractor equations are obtained as :
\begin{eqnarray}
  \label{at1}
  \frac{\partial \F}{\partial v_1} & = & 0
  \,\,\,\Rightarrow 
  \,\,\,1-\frac{v_2}{v_1^2}e^{2\phi}e^2-
  \frac{1}{v_2}e^{2\phi}P^2=0\, ,\\
  \label{at2}
  \frac{\partial \F}{\partial v_2} & = & 0
  \,\,\,\Rightarrow \,\,\,-1+\frac{1}{v_1}e^{2\phi}e^2-
  \frac{v_1}{v^2_2}e^{2\phi}P^2=0\, ,\\
  \label{at3}
  \frac{\partial \F}{\partial \phi} 
  & = & 0\,\,\,\Rightarrow \,\,\,\left( P^2-e^2 \right) =0\, ,\\
  \label{at4}
  \frac{\partial \F}{\partial e} 
  & = & 0\,\,\,\Rightarrow \,\,\, q=\frac{16\pi}{k^2}\left(\frac{v_2}{v_1}e^{2\phi}e \right)\, .
\end{eqnarray}
By combining the first two equations we obtain, $v=v_1=v_2=e^{2\phi}(e^2+P^2)$, which is
also expected from our near horizon geometry analysis as discussed before.  The third
equation gives the value of the moduli at the horizon $e^{-4\phi}=P^2/Q^2$ and therefore,
$v=2PQ$. Now one can check that the entropy is given by the value of the entropy function
$\F$ evaluated at the attractor point:
\begin{equation}
  S=\F=2\pi PQ=\pi v\, .
\end{equation}
Using the electromagnetic field ansatz, one can show that $S=\pi V_{eff}$ and $q=-Q$
(negative sign appears because of our convention for $F_{tr}$).

\subsection{Attractors, dissipation and deep throats}
\label{sec:deep}

Before going on to discuss how the attractor behaviour breaks down for non-extremal black
holes, we consider the analogy between dissipative dynamical systems, their attractor
behaviour and black hole attractors. 

Dissipative dynamical systems are characterized by
the presence of some sort of internal `friction' that tends to contract the phase-space
volume elements. Attractors are states towards which a system starting from certain
initial conditions may evolve after a long enough time. Attractors can be unique states,
called fixed point
attractors.\footnote{For linear dissipative dynamical systems, fixed point attractors are
  the only possible type of attractor. Nonlinear systems, on the other hand, harbor a much
  richer spectrum of attractor types.  For example, in addition to fixed-points, there may
  exist periodic attractors such as limit cycles. There is also an intriguing class of
  chaotic attractors called strange attractors that have a complicated geometric
  structure.}

Now, the extremal black hole attractors sit at the bottom of a infinitely deep $AdS_2$
throat. The authors of \cite{Kallosh:2006bt}, draw an analogy between the radial evolution
of the scalar field down the throat and an under-damped oscillator, which, given
sufficient time will settle to its equilibrium position independent of the initial
conditions. Dissipation erases the ``memory'' of the initial conditions. 

Like any analogy, this one has its uses and short comings. If one examines the
differential equations which describe the radial evolution of the scalars and the metric,
one finds they have to be fine tuned to get the extremal solution. The equations can be
mapped to a mechanical model of a ball rolling up a hill. The radial parameter maps to
time and the initial position and velocity of the ball can be mapped to the asymptotic
values of the scalar and scalar charge. For the attractor black hole solution, the ball
comes to rest precisely at the top. Any trajectory which does not come to rest at the top
corresponds to the scalars blowing up at the horizon which is unphysical.  Clearly one has
to chose the initial velocity of the ball rather judiciously to obtain the solution ---
from the perspective of the mechanical model, this is the complete opposite of attractor
behaviour. However, the point is that this mechanical model does not capture all the
physics of the situation.

Firstly the fine-tuning of the ball's initial velocity, physically corresponds
to the fact that, as we have previously seem, the scalar charge is not an independent
parameter. We should not find this too disturbing since the scalar charge is not a
conserved quantity. 

Secondly, once we consider how our black hole might form we see that it does indeed
display conventional attractor behaviour --- that is the final state does not depend on
initial conditions.  Consider an arbitrary distribution of matter which collapses to form
a blackhole with certain gauge charges. Generically, such a collapse would produce a
non-extremal black hole. This blackhole would then cool via Hawking radiation and approach
extremality.  Since there is only one extremal solution --- the attractor solution --- it
much approach this solution as the black hole cools. As it cools, the throat becomes
deeper and deeper and the black hole becomes more and more sequestered from its
environment eventually completely forgetting about its initial conditions. This is
directly analogous to an under-damped oscillator.  Once we invoke semi-classical effects,
we see that the attractor black hole is attractive in the conventional sense --- given
generic initial conditions (that is some arbitrary matter distribution), we expect to end
up with the attractor solution.

In using this analogy, one should be careful about how one relates radial evolution to
time.  We can associate the radial direction with time evolution, since, as the black hole
cools, the throat becomes deeper.  This should not be confused with the map
between the radial parameter and the time in an artificial mechanical model.

It is amusing to contrast the fine tuning of the mechanical model with the genericity of
the attractor solution we argue for above. Perhaps this fine tuning corresponds to the
fine tuning that would be required to form an extremal black hole without invoking
semi-classical effects.

Finally, we note that there may be flat directions in the entropy function. This can lead
to generalised attractor behaviour in which the entropy is independent of the moduli but
the near horizon geometry is not \cite{Astefanesei:2006dd}. Extending our analogy, one
might suppose that flat directions correspond to modes that do not couple to the
dissipation and can persist even in the extremal limit. The black hole forgets enough to
ensure the entropy does not depend on the moduli.
\footnote{ Notwithstanding information leaking through the deep throat, on might say that
  the generalised black hole attractor behaves like an ideal `tricky' politician, choosing
  to forget (or erase) only compromising details.}

\subsection{Non-extremal solutions and unattractor equation}\label{sec:non-extr-solut}

We would like to understand the relation between the entropy and the value of effective
potential at the horizon for the non-extremal black holes. The first observation is that
the near-horizon geometry of a non-extremal black hole (\ref{nonextreme}) does not contain
an $AdS_2$ part. Since, as we have seen, the $AdS_2$ symmetries implied the attractor
behaviour it is plausible to suppose that the converse applies --- in the absence of these
symmetries there is no attractor behaviour. Then, the black hole horizon is not an
attractor for the moduli. The  effective potential evaluated at the horizon and the entropy
will receive corrections away from the attractor value which depend on the asymptotic values
of the moduli.

We investigate (\ref{constraint}) and (\ref{eq3}) at the horizon. Using  some results 
from subsection \ref{efunction}, namely,
\begin{equation}
  f_0=4\pi T, \qquad \epsilon=\frac{f_0S}{\pi}=4TS=2c\,. 
\end{equation}
we try to  write things in terms of the temperature and entropy as much as possible.
Equation, (\ref{constraint}) gives us a relation between the entropy and the value of the
effective potential at the horizon:
\begin{equation}
  \label{misto}
  V_{eff}=\frac{\epsilon}{2 f_0}\left(1+\frac{\epsilon f_1}{f_0}\right)
  = \frac{S}{2\pi}\left( 1 + \frac{S f_1}{\pi}\right)\, .
\end{equation}
The other equation, (\ref{eq3}), evaluated at the horizon gives us
\begin{equation}
\label{misto2}
  \frac{\partial V_{eff}}{\partial \phi}=
  \frac{S^2}{\pi^2}\sqrt{2f_0f_2+\frac{1}{2}(\frac{\pi}{S}-f_1)(\frac{\pi}{S}+3f_1)}\,\, .
\end{equation}

There is a class of near-extremal black holes which break supersymmetry, but whose entropy
can still be accounted for by microscopic counting \cite{Horowitz:1996fn, Callan:1996dv}.
These are 5-dimensional black hole solutions and in the extremal limit, the near-horizon
geometry contains an $AdS_3$ factor, rather than an $AdS_2$.  In this case one can use the
$AdS_3/CFT_2$ correspondence and the Cardy formula to compute the entropy. Unfortunately,
there is no entropy function formalism for this kind of black holes.  However, it was
pointed out in \cite{Sahoo:2006vz} that there is a nice relation between $AdS_3$ and
$AdS_2$ in the context of attractor mechanism. Also, Maldacena \cite{Maldacena:1996iz}
observed that the supersymmetry of the theory describing the excitations of the D-branes
is similar to ${\cal N}=2$ in four dimensions, the supersymmetry we are interested in.
That is an $(1+1)$-dimensional field theory with $(4,4)$ susy --- this is the susy
left unbroken by the extremal D-branes. There are vector multiplets and hypermultiplets
and the distinction between them is that they have different transformation properties
under $R$ symmetries.

Let us comment on the role of the effective potential in the case of near-extremal black
holes. We saw that, in the extremal case, the fixed $t$-surface takes the geometry of an
infinite cylinder (the `infinite throat'). It seems that the horizon has been pushed away
to infinity, though one can still fall into the black hole in finite proper time since the
horizon is still a finite distance away in time-like or null directions. The near horizon
geometry of the non-extremal black hole is rather similar to Rindler space as opposed to
$AdS_2\times S^2$. In this case, the effective potential is not generically extremised and
the attractor behaviour is absent.  However, there is a special case when the attractor
is still useful. Let us consider black holes with large charges ($Q\gg 1$). Now, let us
repeat the arguments of \cite{Maldacena:1996iz} to explain the microscopic/macroscopic
agreement for the near-extremal black holes. In our discussion, we can keep $g_s$ small
(closed string effects are small) and we obtain a strong coupling regime because of the
large number $Q$ of branes: the fundamental strings couple weakly to each other but
interact strongly with the collection of D-branes.  The effective open string coupling is
$g_sQ$. For $g_sQ\ll 1$ we obtain the domain of validity of the D-brane perturbation
theory and for $g_sQ\gg 1$ we obtain the semi-classical black hole domain.

Using our equation (\ref{misto}), one finds that, in the near-extremal limit and for large
charges, the entropy is still given by the value of the effective potential at the
horizon. In the near-extremal limit, the effective potential depends on the values of
moduli at infinity. However, the string coupling is small and so the corrections
received by the effective potential are small in comparison with its value in the extremal
limit.  The near-horizon geometry is approximately $AdS_2\times S^2$ and the attractor
mechanism still works in this case.

\subsection{First order perturbation analysis of non-extremal black
  holes}\label{sec:first-order-pert}

To put our discussion of non-extremal unattractive black holes on a more quantitative
footing, we study the effect of perturbing a non-extremal black hole away from the
attractor point.  We will start with a non-extremal Reissner-Nordstrom black hole where
the scalar field is fixed at the attractor value everywhere. We ask how the solution
changes if we shift the asymptotic value of the scalar slightly away from the attractor
point. In particular we are interested in how the value of the scalar field changes at the
horizon. This extends the perturbation analysis of \cite{Goldstein:2005hq} which
mainly considered extremal black
holes.\footnote{While the non-extremal case was discussed in \cite{Goldstein:2005hq}, the
  first order analysis was only done approximately. Our results confirm the qualitative
  picture discussed there. }

For concreteness we consider an effective potential of the form 
\begin{equation}
  V_{eff}= e^{-\alpha \phi}Q^2 + e^{\alpha \phi}P^2.
\end{equation}

We can consistently set the scalar, $\phi$, to a constant if it is at the attractor point,
$\p_\phi V_{eff}=0$. Extremising the effective potential one finds
\begin{equation}
  e^{2\alpha\phi_{\attr} }=Q^2/P^2
\end{equation}
where $\phi_{\attr}$ is the attractor value.  Now, with the scalar constant everywhere, we
effectively have Einstein-Maxwell gravity which in particular has the non-extremal
Reissner-Nordstrom black hole as a solution.

We now perturb the non-extremal Reissner-Nordstrom black hole by assuming that the scalar
varies slightly from the attractor value:
\begin{equation}\label{scalar.pert}
  \phi(r)=\phi_\attr+\varepsilon \phi_1(r)
\end{equation} 
Here $\varepsilon$ is a small parameter we use to organise the perturbation theory. As
discussed in \cite{Goldstein:2005hq}, the scalar perturbation (\ref{scalar.pert}), will
source second order perturbations to the metric. This back-reaction will in turn source
second order perturbations to the scalar field etc. 

Let us write
\begin{eqnarray}
  a^2&=&a_0^2+\varepsilon^2a_2  \\
  b&=&b_0+\varepsilon^2b_2\nonumber
\end{eqnarray}
where $b_0$ and $a_0$ are  the unperturbed non-extremal Reissner-Nordstrom solution.
They are given by,
\begin{eqnarray}
  a_0^2&=&\frac{1}{r^2}(r-r_+)(r-r_-)\\
  b_0&=& r
\end{eqnarray}
with $r_\pm=M\pm\sqrt{M^2-(\bar{Q}^2+\bar{P}^2)}=M\pm\sqrt{M^2-2|QP|}$.

Now, expanding the equation for the scalar field,~(\ref{dilaton}), to first order in
$\epsilon$ in the non-extremal Reissner-Nordstrom background, we find
\begin{equation}
  \p_r\left[ (r-r_+)(r-r_-)\p_r\phi_1\right]=\frac{\beta^2\phi_1 }{r^2} +{\cal O}(\varepsilon)
\end{equation}
where $\beta^2=\half\p^2_\phi V_{eff}(\phi)|_{\phi=\phi_0}=\alpha^2|QP|$.

It is convenient to define a new variable, $x$, given by
\begin{equation}
  x=\frac{r_+(r-r_-)+r_-(r-r_+)}{r\left(r_+-r_-\right)}
   =(r_+-r_-)\frac{r_+e^{-c\tau}+r_-e^{c\tau}}{e^{c\tau}-e^{-c\tau}},
\end{equation}
for which the inner and outer horizons are at  $x_\pm=\pm1$  and $r=\infty$
is at  $x_\infty=(r_++r_-)/(r_+-r_-)$.

In terms of $x$, the equation for $\phi_1$ is just Legendre's equation: 
\begin{equation}
  \p_x((x^2-1)\p_x\phi_1)={\textstyle {\half}}\alpha^2 \phi_1
\end{equation}
which has the solution
\begin{equation}
  \phi_1=c_1P_\gamma(x)
\end{equation}
where $\gamma=\half(\sqrt{1+2\alpha^2}-1)$ and $P_\gamma$ is a Legendre function of the
first
kind.\footnote{The other linearly independent solution is a Legendre function of the
  second kind which diverges at the horizon.}
When $\gamma$ is an integer, $P_\gamma$ is a polynomial.  It is amusing notice that this
is the same $\gamma$ we used to categorise the exact solutions. Perhaps in these cases the
perturbations series can be exactly summed.

To first order in $\varepsilon$, we find
\begin{equation}
  \phi(x)=\phi_\attr+(\phi_\infty-\phi_\attr)\frac{P_\gamma(x)}{P_\gamma(x_\infty)}
\end{equation}
In particular, deviation of the scalar field at the horizon from the attractor is just
proportional to its deviation at infinity. Now, $x_\infty$ is inversely proportional to
the temperature, so when the temperature is small, $x_\infty$ is large. Using the
asymptotic form of the Legendre function \cite{Morse:1953:MTP},
$P_\gamma(x_\infty)\sim x_\infty^\gamma\sim T^{-\gamma}$, we find that the deviation of the
scalar field from the attractor value at the horizon goes like the deviation at infinity
times the temperature to some power:
\begin{equation}
  (\phi_{Horizon}-\phi_{attractor})\sim(\phi_\infty-\phi_{attractor}) T^\gamma.
\end{equation}
This formula holds for small deviations from the attractor values and small
temperatures. Notice that in the limit $T\rightarrow0$ we recover the attractor behaviour
and for near extremal black holes we have approximate attractor behaviour as expected. 

Now we return to (4.38) and (4.39). Since the change to the metric are second order in
$\varepsilon$, so are the changes of the coefficients $f_i$. We see that $V_{eff}$ remains
unchanged to first order but, due to the square root on the right hand side of (4.39),
$\p_\phi V_{eff}$ changes to first order. So, to first order the scalar is governed by the
same effective potential but it does not sit at the minimum at the horizon. In other
words, for the non-extremal black hole, the infinite $AdS$ throat has been capped off and the
scalar field doesn't have ``time'' to reach the minimum of its potential before it hits
the horizon.

Although the scalar field is shifted, it does not necessarily follow that the entropy
changes.  Since changes to the metric appear at second order, the entropy actually remains
unchanged to first order. However, we expect moduli dependence of the entropy to appear at second
order. Since, in general, the second order perturbation analysis is rather complicated, we
examined some of the exact solutions and found that moduli
dependent corrections to the entropy do indeed appear at second order.

\subsection{(Un)attractors and special geometry}
\label{sec:sg}

The Lagrangian in (\ref{actiongen}) can be embedded in ${\cal N}=2$ supergravity theory for
certain special values of the couplings.  In this subsection, we briefly review the
analysis of the (un)attractor equations in ${\cal N} =2$ special geometry language
\cite{Ferrara:1997tw, Kallosh:2006bt} and show the relation with our results.

The bosonic part of the ${\cal N}=2$ supergravity action coupled to arbitrary number of vector
multiplets is given
by\footnote{For details on special geometry formulation, we refer to 
\cite{deWit:1984pk, deWit:1984px}.}
\begin{eqnarray} - \frac{R}{2} + G_{a\bar a} \partial_{\mu} z^a \partial_{\nu} \bar
  z^{\bar a} + Im {\cal N}_{\Lambda\Sigma}{\cal F}^{\Lambda}_{\mu\nu}
  {\cal F}^{\Sigma}_{\lambda\rho}g^{\mu\lambda} g^{\nu\rho} + Re
  {\cal N}_{\Lambda\Sigma}{\cal F}^{\Lambda}_{\mu\nu} \star
  {\cal F}^{\Sigma}_{\lambda\rho} g^{\mu\lambda}g^{\nu\rho}.
\end{eqnarray}
Here $G_{a\bar a}$ is the metric of the scalar manifold and $Re{\cal N}$ and $Im{\cal N}$
components of ${\cal N}$ are negative definite scalar dependent vector couplings.  Their
explicit expressions can be obtained in terms of the symplectic sections of the underlying
${\cal N}=2$ theory and from the prepotential.  The negative of the real and imaginary part of
the vector couplings in the above ${\cal N}=2$ action can be schematically identified with our
previous quantities $\tilde f_{ab}$ and $f_{ab}$ respectively.
                                                                                
One can construct the symplectic invariant quantity
$Z(z, \bar z, p, q)|^2 + |D_aZ(z, \bar z, p, q)|^2$ which can be identified with the
scalar dependent effective potential $V_{eff}$.  Here, $Z$ is the central charge in ${\cal N}=2$
supergravity theory and $D_a Z$ is the K\"ahler covariant derivative, $z$ are the complex
moduli, $p$ and $q$ are the magnetic and electric charges respectively. The central charge
is given by the expression
\begin{eqnarray} Z(z, \bar z, q, p) = e^{\frac{K(z, \bar z}{2}} \left ( X^{\Lambda}(z)
    q_{\Lambda} - F_{\Lambda} p^{\Lambda} \right )\, ,
\end{eqnarray}
where $K(z, \bar z)$ is the K\"ahler potential.  So the effective potential is given in a
simple form:
\begin{eqnarray} V_{eff}(z, \bar z, q, p) = |Z(z, \bar z, p, q)|^2 + |D_aZ(z, \bar z, p,
  q)|^2\, .
\end{eqnarray}
                                                                                 
The above form of $V_{eff}$ can be simplified to obtain an expression in terms of
electric, magnetic charges as well as the real and imaginary part of the vector coupling
${\cal N}$:
\begin{eqnarray} V_{eff} = - \frac{1}{2} \left(\begin{array}{cc} p & q
    \end{array} \right) 
  \left(\begin{array}{cc}
      Im {\cal N} + Re {\cal N} Im {\cal N}^{-1} Re {\cal N}  
      & - Re {\cal N} Im {\cal N}^{-1}
      \\
      - Im {\cal N}^{-1} Re {\cal N} & Im {\cal N}^{-1}
    \end{array} \right) \left(\begin{array}{c}
      p \\
      q
    \end{array} \right) \;,
  \label{matrixM}
\end{eqnarray}
where, we have suppressed the indices $I, J$.  This is equivalent to the expression for
the effective potential obtained in \cite{Goldstein:2005hq} which has been derived by
using the metric ansatz and the equations of motion.
                                                                                 
The metric of the spherically symmetric solution is given by
\begin{eqnarray} ds^2 = e^{2U} dt^2 - e^{-2U} \left [ \frac{c^4}{\sinh^4c\tau} d\tau^2 +
    \frac{c^2}{\sinh^2 c\tau} d\Omega^2 \right ] \end{eqnarray} Then the constraint
becomes,
\begin{eqnarray} \left (\frac{dU}{d\tau} \right )^2 + \left | \frac{dz}{d\tau} \right |^2
  + e^{2U} \left (|Z(z, \bar z, q, p)|^2 + |D_a Z(z, \bar z, q,p)|^2 \right ) = c^2
\end{eqnarray}

The constraint expression evaluated at infinity (at
$\tau \rightarrow 0, U \rightarrow M\tau$) is given by,
\begin{eqnarray} M^2 (z_{\infty}, \bar z_{\infty}, p, q) - |Z(z_{\infty}, \bar z_{\infty},
  p, q)|^2 = c^2 + |D_aZ(z_{\infty}, \bar z_{\infty}, p, q)|^2 - G_{a\bar a} \Sigma^a
  \bar\Sigma^{\bar a}
\end{eqnarray}
For BPS configuration,
\begin{eqnarray}
  M^2(z_{\infty}\bar z_{\infty}, p, q) = |Z(z_{\infty}, \bar z_{\infty}, p, q)|^2, \qquad
  c = 0, \qquad G^{a\bar a}\bar D_{\bar a}Z(z_{\infty}, \bar z_{\infty}, p, q) =
  \Sigma^a 
\end{eqnarray}
                                                                                 
For extremal solution, $c^2 = 2 S T = 0$, (here we have used a different normalization for
the parameter $c$ as compared to our previous discussion) but when
$D_a Z(z, \bar z, p,q) \neq 0$, this describes non BPS solutions.  For non-extremal
solutions $c^2 \neq 0$ and $D_a Z(z, \bar z, p, q) \neq 0$.
                                                                                 
The condition for the attractor is obtained by knowing the critical point of the effective
potential. Using special geometry identities, the critical point of $V_{eff}$ is given by
the expression,
\begin{eqnarray}
  \partial_a V_{eff} = 2 (D_a Z)\bar Z + i C_{abc} G^{b\bar m} G^{c\bar n} \bar
  D_{\bar m}\bar Z \bar D_{\bar n} \bar Z \end{eqnarray}
This shows that {\it l.h.s.} is zero when
$D_a Z = \bar D_{\bar a}\bar Z = 0$ which means that the critical point of $V_{eff}$
coincides with the critical point of the central charge.
                                                                                 
The second condition for the existence of an attractor is obtained by evaluating the
second derivative of the effective potential at the critical point. Using special geometry
identities, one gets,
\begin{eqnarray} \left ( \bar D_{\bar a} D_b V_{eff} \right )_{cr} = \left (\bar\partial
    _{\bar a}
    \partial_b V_{eff} \right )_{cr} = 2 G_{\bar a b} V_{eff}(cr) 
\end{eqnarray}
which shows that the sign of the second derivative of the effective potential is positive
when the sign of the moduli space metric at the critical point is positive.
                                                                                 
Equivalently, one can also obtain these conditions by considering the equation of motion
for the scalars and assuming the moduli space to be a complex K\"ahler manifold with a
K\"ahler metric $G_{a\bar b}$.
                                                                                 
The equation of motion for the scalar is given by \cite{Kallosh:2006bt},
\begin{eqnarray}
  \partial_{\tau}(\partial_{\tau}z^a) + \Gamma^a_{bc}(z, \bar z)
  \partial_{\tau} z^b \partial_{\tau} z^c = G^{a\bar b} e^{2U} \frac{\partial
    V}{\partial \bar z^{\bar b}} 
\end{eqnarray}
                                                                                 
For extremal (BPS and non BPS) solution, the l.h.s. at the horizon is zero, so one gets
the condition
\begin{eqnarray}
  \left.\frac{\partial V}{\partial z^a}\right |_{z^a_h} =  0 
\end{eqnarray}
where, $z^a_h$ is the value of the scalar field at the horizon.

For non-extremal case, the attractor equation reduces to (after some change of
coordinates),
\begin{equation}
  \label{eq:attr:ne}
  \left. (z^a)^{''}\right |_{\rho = 0} = g^{a\bar b} 
  \left. \frac{\partial V}{\partial \bar z^{\bar b}}\right |_{z^a_h}.
\end{equation}
Here, $\rho \rightarrow 0$ is the near horizon limit and $\rho$ is related to $\tau$ as
$\rho = 2 e^{c\tau}$.  The metric function in the new coordinate is given as,
$e^{2U} \rightarrow (-c^2\rho^2/r_h^2)$. The near horizon geometry is then given as :
\begin{equation}
  \label{eq:ne:metric}
  ds^2 = \rho^2 dt^2 - (r_h)^2 d\rho^2 - (r_h)^2 d\Omega^2, ~~~~~~ \rho \rightarrow 0   
\end{equation}
Certainly, the {\it l.h.s.} of the attractor equation,~(\ref{eq:attr:ne}), is not zero and
hence the derivative of the effective potential is not zero and hence there is no
attractor phenomenon.  This is reflected in our equation (\ref{misto2}) which
has been derived from the scalar field equation of motion and using the horizon values of
the derivative of the moduli as well as $a^2$, $b^2$ appearing in the metric function.  We
have shown there that the {\it r.h.s.} of equation (\ref{misto2}) is not equal to zero. This
shows that the attractor phenomenon does not occur in the non-extremal case.

\section{The role of non-supersymmetric attractor in 
microscopic/macroscopic 
entropy matching}  \label{sec:entropy.match}

The extremality condition was enough to constrain the near-horizon geometry and ensure
that the entropy is independent of the asymptotic values of the moduli --- the entropy only
depends on the modulus independent product $\bar{P}\bar{Q}=PQ$. In this
section\footnote{DA would like to thank Ashoke Sen for discussions on this section.}
we argue that the attractor mechanism is at the basis of the matching between
microscopic and macroscopic entropy of certain extremal non-BPS black holes. In
particular, we consider the examples as discussed in
\cite{Kaplan:1996ev,Emparan:2006it}. 

The thermodynamics of extremal black holes is very tricky (see, also, the Discussion
section). In many cases, the horizon area is finite and one expects the entropy to be
non-zero.  Therefore, a vanishing entropy on the Euclidean section should prevent us from
trusting the Euclidean semi-classical calculations.  A strong argument to support a
non-vanishing entropy for extremal black holes comes from string theory which provides a
nice microscopic interpretation. In string theory, the entropy of an extremal BPS black
hole is computed by counting the degeneracy of D-branes states --- D-branes are the
constituents from which the black hole is formed.  That is equivalent to counting the BPS
states (lowest mass states at fixed charges) in the D-brane world-volume theory. Supersymmetry
is at the basis of the non-renormalization theorems that ensures that the ground state
degeneracy is a kinematic quantity rather than a dynamical one (it is independent of the
strength of the string coupling). Then, the counting of the number of D-branes at weak
coupling agrees with the classical area law of the black hole at strong coupling. 

The large
\footnote{Large black holes are the black holes for which the spacetime curvature is weak
  outside the horizon --- of course when the curvature is strong, e.g. is blowing up at
  the singularity, the stringy effects are important.}
non-susy black holes share an important property with their BPS cousins: they have the
lowest possible mass in the quantum theory (due to the extremality condition) and there is
no other black hole state to which they can decay by Hawking radiation. Then, their
temperature should vanish.  The extremality condition acts as the cosmic censorship
preventing a minimum mass black hole to decay in a naked singularity. An important
question arises here, namely, is there any D-brane microscopic configuration to describe
such a non-BPS black hole?  The answer is affirmative and in what follows, we present a
concrete example.

In \cite{Emparan:2006it}, an intriguing example of microstates counting for a neutral
black hole has been proposed that precisely reproduces the Hawking-Bekenstein entropy. The
non-rotating case corresponds to our solution II. This solution can be embedded in string
theory \cite{Larsen:1999pp} by identifying the KK circle with the M-theory circle. In
this way, the magnetic charge becomes the charge of the D-6 brane and the electric charge
is the D-0 brane charge.

We will discuss this case using the effective potential formalism, but the computations
using the entropy function are very similar with the calculations in the section
\ref{efunction}. The effective potential is given by
\begin{equation}
  V_{eff}=e^{\alpha\phi}(Q)^{2}+e^{-\alpha\phi}(P)^{2}\, ,
  \label{seffpot2b}
\end{equation}
with $\alpha=2\sqrt{3}$. Then, the value of the modulus at the horizon and the
entropy are
\begin{equation}
  e^{4\sqrt{3}\phi_H}=\frac{P^2}{Q^2}\, , \,\,\,\,\,\,\,\,\,\,\,\,\,\, S=\pi V_{eff}(\phi_H)=2\pi PQ.
\end{equation}
These expressions involve only the charge parameters, the moduli cancel out.  This is
interpreted as a signal that a clear connection to the microscopic theory is possible.
This black hole is not supersymmetric and this is in agreement with the absence of the
bound states of $D_0$ and $D_6$ branes. However, in \cite{Emparan:2006it}, a simple string
description based on non-supersymmetric, quadratically stable, $D_0-D_6$ bound states
\cite{Taylor:1997ay} was provided --- this statistical prescription precisely reproduces
the Hawking-Bekenstein entropy.

The electric and magnetic charges are quantized according to Dirac's quantization rule and
we obtain
\begin{equation} {2PQ\over G_{4}}= n_{Q}n_{P}\, ,
\end{equation}
where the integers $n_{Q}$ and $n_{P}$ can be interpreted after the embedding in M-theory
as the number of $D0$- and $D6$-branes, respectively. The entropy of extremal black hole
can be rewritten \cite{Larsen:1999pp} as:
\begin{equation}
  S = \pi n_{P}n_Q = \pi N_0N_6\, .
\end{equation}
 
The excitations of the D-branes system at low energies are described in terms of a moduli
space approximation. Since our black hole is non-BPS, one can not use the
non-renormalization theorems to argue that low energy theory does not receive corrections
when the string coupling is increased. Therefore, the black hole dynamics in the strong
coupling regime is described by a different moduli space.  Now, it is important to
remember that the near-horizon geometry of the non-BPS extremal black holes is the same 
as for the BPS extremal black holes, namely $AdS_2\times S^2$. Then
even if we start with different moduli spaces (in different regimes), the moduli are
attracted to the horizon to the same values. Then, this seems to be the reason for the
mysterious microscopic/macroscopic entropy match in
\cite{Emparan:2006it}.

We will comment more on the validity of our proposal in the Discussion section.

\section{Discussion}\label{sec:discussion}

In classical general relativity, no hair theorems impose strong constraints on the
possibility of obtaining solutions of the Einstein equations coupled to non-trivial scalar
fields. A crucial ingredient for their proof is that the scalars be minimally coupled to
gravity and other fields. In this paper we have considered black holes with scalar fields
which are non-minimally coupled to gauge fields. Clearly, this is a concrete possibility
for evading no hair theorems. Indeed, we have seen how the non-zero asymptotic scalar
charges and the values of moduli at infinity play a role in the first law of
thermodynamics. However, this is not considered as a drastic violation of the no hair
theorems \cite{Coleman:1991jf, Shapere:1991ta, Coleman:1991ku}. The reason is that the
scalar charges are not independent parameters, but are given functions
of the other
asymptotic charges which characterise the solution.

Let us consider, for example, the part of the Lagrangian containing the moduli and the
moduli coupled to gauge fields.  The Lagrangian reduces to the standard Einstein-Maxwell
form if the moduli are constant. However if $F_{\mu\nu}F^{\mu\nu}\neq 0$ and the scalar
field is not at the minimum of its effective potential, the field equation for $\phi$ will
not be satisfied by taking $\phi$ to be a constant.

This means that the non-vanishing electromagnetic field can also be understood as a source
for the moduli. As a result, the scalar charges have been called {\it secondary hair} by
the authors of \cite{Coleman:1991jf, Shapere:1991ta, Coleman:1991ku}.\footnote
{We prefer the term stubble.}  They are generated because the basic fields (associated
with mass, angular momentum, and gauge charges) act also as sources for the moduli. This
should be contrasted with {\it primary hair} which would be due an asymptotic scalar
charge which is completely independent of the other charges.  It is significant that they
do not represent a new quantum number associated with the black hole. However, in string
theory the scalar fields, referred to as {\it moduli} are interpreted as local coupling
constants rather than matter fields and the notion of scalar charge is somehow
misleading. While conventional in general relativity, one would not normally consider
variations of the moduli at infinity, there are case in both string theory and general
relativity where one might be lead to consider them hence introducing a new term in the
first law (\ref{1law}). For example, one may be interested in time-dependent cosmological
situations in which $\phi_{\infty}$ becomes
dynamical\footnote{Since the derivation assumes asymptotically flat space, one would 
  require sufficiently flat background for the analysis to remain valid.},
one may wish to understand the behaviour of the black holes under slow adiabatic changes
of $\phi_{\infty}$ or one may wish to compare black holes at different points in moduli
space.

The non-extremal black holes have a non-zero temperature that can be evaluated by
eliminating the conical singularity in the Euclidean section. Then, the Euclidean geometry
becomes a `cigar' and so the Euclidean time circle closes off smoothly.  On the other
hand, for an extremal Euclidean black hole the topology changes. The Euclidean time circle
does {\it not} close off and so there is no conical singularity. In this case, one is
forced, either to work with an arbitrary periodicity of the Euclidean time leading to
ambiguous results, or simply to ignore the Euclidean time method.  However, in the
Lorentzian section the picture is quite satisfactory: an extremal black hole is obtained
by continuously sending the surface gravity of a non-extremal black hole to zero. While the
surface gravity (i.e. the temperature) vanishes,  the area of the horizon
(i.e. the entropy) can remain finite.  These results strongly suggest that the entropy of
an extremal black hole with a non-vanishing horizon area is non-zero.

The extremal supersymmetric black holes play a central role in providing a statistical
foundation for black hole thermodynamics in string theory.  In all known cases,
supersymmetric (static) black holes are also extreme black holes --- the converse is not
true. This can be understood by the fact that the BPS black holes are stable systems
corresponding to the lowest possible mass in the quantum theory and should not
radiate. Then, their temperature should vanish and so they are extremal. On the other
hand, not all extremal black holes saturate BPS bounds and they can break supersymmetry
--- the mismatch between the extremality and BPS conditions is quite general
\cite{Ortin:1996bz}. There are two kinds of extremal non-susy solutions. The first one
contains extremal black holes which can not be embedded in supergravity theories (e.g., a
subset of the solutions Ib). The second one contains extremal black holes which are
solutions of supergravity theories but are not supersymmetric (e.g., the solution II).

It was discovered long time ago that, in four-dimensional, ungauged ${\cal N}=2$ supergravity,
the BPS black hole solutions exhibit fixed-point attractor behavior near the
horizon. However, recently it was understood that the near-horizon extremal geometry
\cite{Sen:2005wa} is at the basis of the attractor mechanism,
rather than supersymmetry. It is just more convenient to solve the supersymmetry
transformations for the gravitino and gauginos in a bosonic background of (${\cal N}=2$)
supergravity --- these transformations depend linearly on the first
derivatives. Consequently, to find BPS black hole solutions one has to solve first order
differential equations (the attractor equations near-horizon become algebraic)
\cite{Sabra:1997kq, Sabra:1997dh}.

When the BPS bound is saturated, the entropy is determined microscopically just by the
charges. However, the charges are quantized and then the entropy should 
also be a discrete
quantity. Instead, the moduli are continuous parameters of the internal manifold. For
consistency with the discreteness of the entropy, the values of the moduli at the horizon
can not have any arbitrary values. The attractor mechanism provide an explanation for why
the moduli are fixed at the horizon. However, this argument is not based at all on
supersymmetry and this was one important reason for investigating non-supersymmetric
attractors in \cite{Goldstein:2005hq}.

In the previous section, we proposed that the attractor mechanism is at the origin of the
microscopic/macroscopic match of some non-BPS extremal black holes
\cite{Emparan:2006it, Kaplan:1996ev} (see, also, \cite{Dabholkar:1997rk} for a five-dimensional example). 
We used the effective potential method to show that the Kaluza-Klein horizon is an 
attractor for the moduli and have explicitly shown that there is just one minimum. This is the explanation 
for why the entropy is independent of continuous parameters (coupling constants). It is also worth to be 
mentioned that the existence of just one minimum imply that  there can not 
be jumps in entropy moving from weak coupling to strong coupling coupling. 
Although this proposal is certainly suggestive, one
should take it with some caution --- similar arguments are not valid when the basin of
attraction is not unique. In general, the effective potential method will provide more
information about the attractor behavior than the entropy function.  For example, if there
is more than one attractor fixed point, then a study of the effective potential will make
it clear which minimum can be obtained by starting with different boundary conditions. 
In this case, our arguments fail, because by changing the coupling $g_sQ$, the moduli can 
end up in a different domain of attraction and the value of the entropy will
change.\footnote{The near-horizon geometry remains $AdS_2\times S^2$ even after adding $\alpha'$
  corrections --- the radia of $AdS_2$ and $S^2$ receive corrections, but the geometry
  does not change.} For the KK solution we have found that the effective potential has just one minimum and we
expect this is also true for the black holes of \cite{Kaplan:1996ev}.  Another point worth
to be mentioned is that, for consistency with the macroscopic picture, the microscopic
configuration of branes should be also non-supersymmetric but stable. For the case at hand
--- KK black hole --- it is known \cite{Taylor:1997ay} that, indeed, this is true.  The
0-brane and the 6-brane repel one another and so, in general, a point-like 0-brane placed
on or near a 6-brane gives rise to a non-susy configuration. However, 
a D0-D6 brane configuration has been proposed in
\cite{Taylor:1997ay} which satisfies the classical
equations of motion and is {\it classically stable} to quadratic order. 
That is, a
set of four 0-branes which are smeared out over four 6-branes wrapped on a six-torus ---
this configuration served as a basis of the microscopic picture in
\cite{Emparan:2006it}. These metastable states are interpreted as some kind of long-lived
resonances composed of 0-branes and 6-branes and so the microscopic and the macroscopic
pictures are consistent with one another. One more puzzle is related to the lack of
non-renormalization theorems for the extremal non-BPS black holes.  In the strong coupling
regime the extremal black hole can still be thought of as the black hole with the lowest
mass.  However, by changing the coupling, the mass will receive corrections and the
statistical entropy definition should be revised.

The counting in \cite{Emparan:2006it} requires $N_6\gg 4$, but the configurations
constructed in \cite{Taylor:1997ay} can be found even for small numbers of branes. It will
also be interesting to reproduce the entropy of KK black hole in this case. Our
investigation in subsection \ref{sec:non-extr-solut} strongly suggests that, due to the
attractor mechanism, a computation of the near-extremal KK black hole entropy with large
charges should be also possible.

The rotating case was also studied in \cite{Astefanesei:2006dd} by using the entropy
function. It is worth noticing that the long throat of $AdS_2$ is also present in the
near-horizon geometry of the extremal rotating black hole. Unfortunately, in the rotating
case, it is difficult to construct an effective potential when the moduli are not
constants. It will be interesting to find an effective potential analogous to (\ref{misto}) 
and study the near-extremal rotating black
holes. It will also be interesting to investigate the thermodynamics of the non-extremal
black holes by using the `counter-term' method developed in
\cite{Astefanesei:2005ad, Mann:2005yr, Astefanesei:2006zd}.

\section*{Acknowledgements}
 The authors would like to thank Ashoke Sen for suggestions, useful
discussions, and sharing with them some of his related ideas.  We would also like to thank
Sandip Trivedi for interesting conversations.  DA would like to thank the members of HRI
string group for their feedback during the discussions at String lunch. The work of
D.A. and K.G. is supported by the Department of Atomic Energy, Government of India.

\appendix

\section{Effective potential}
\label{appendix1}
Let us derive the expression of the effective potential, $V_{eff}$, by rewriting
(\ref{dilaton}) as (\ref{eq3}).  We have
\begin{equation}
  \partial_{\mu}(\sqrt{-G}g_{ij}G^{\mu\nu}\partial_{\nu}\phi^j)
  =\sqrt{-G} \left[\frac{1}{4} \frac{\partial f_{AB}}
    {\partial \phi^i} F^A _{\phantom{A}\mu\nu} F^{B\, \mu\nu} 
    +\frac{1}{8}\frac{\partial \tilde f_{AB}}{\partial \phi^i} 
    F^A_{\phantom{A}\mu\nu} F^B_{\phantom{B}\rho \sigma} 
    \epsilon^{\mu\nu\rho\sigma}\right],
\end{equation}
with $\sqrt{-G}=b^2\sin\theta$. Then
\begin{equation}
  \partial_{r}(b^2\sin\theta g_{ij}G^{rr}\partial_{r}\phi^j)
  =(b^2\sin\theta) \left[\frac{1}{4} \frac{\partial f_{AB}}
    {\partial \phi^i} F^A _{\phantom{A}\mu\nu} F^{B\, \mu\nu} 
    +\frac{1}{8}\frac{\partial \tilde f_{AB}}{\partial \phi^i} 
    F^A_{\phantom{A}\mu\nu} F^B_{\phantom{B}\rho \sigma} 
    \epsilon^{\mu\nu\rho\sigma}\right],
\end{equation}
and
\begin{equation}
  \partial_{r}(b^2a^2g_{ij}\partial_{r}\phi^j)
  =b^2\left[\frac{1}{4} \frac{\partial f_{AB}}{\partial \phi^i} 
    F^A _{\phantom{A}\mu\nu} F^{B\, \mu\nu} 
    +\frac{1}{8}\frac{\partial \tilde f_{AB}}{\partial \phi^i} 
    F^A_{\phantom{A}\mu\nu} F^B_{\phantom{B}\rho \sigma} 
    \epsilon^{\mu\nu\rho\sigma}\right] .
\end{equation}
Now, we calculate $V_{eff}$ by using its definition (\ref{eq3}):

\begin{equation}
  \label{calcul}
  \frac{\partial V_{eff}}{\partial \phi^i}=2b^4\left[\frac{1}{4} 
    \frac{\partial f_{AB}}{\partial \phi^i} F^A _{\phantom{A}\mu\nu} 
    F^{B\, \mu\nu} 
    +\frac{1}{8}\frac{\partial \tilde f_{AB}}{\partial \phi^i} 
    F^A_{\phantom{A}\mu\nu} F^B_{\phantom{B}\rho \sigma} 
    \epsilon^{\mu\nu\rho\sigma}\right].
\end{equation}

To get (\ref{defpotgen}), we need to use (\ref{fstrenghtgen})

\begin{equation}
  F^A=f^{AB}(Q_{B}-{\tilde f}_{BC}P^C) {1\over b^2} dt\wedge dr + 
  P^A \sin \theta  d\theta \wedge d\phi=F^A_{tr}dt\wedge dr+
  F^A_{\theta\phi}d\theta \wedge d\phi,
\end{equation}

and so we find:

\begin{equation}
  \frac{1}{4} \frac{\partial f_{AB}}{\partial \phi^i}F^A _{\phantom{A}\mu\nu} 
  F^{B\, \mu\nu}=\frac{1}{4}2\frac{\partial f_{AB}}{\partial \phi^i}
  (G^{rr}G^{tt}F^A_{tr}F^B_{tr} + G^{\theta\theta}G^{\phi\phi}F^A_{\theta\phi}
  F^B_{\theta\phi})
\end{equation}

\begin{equation}
  \nonumber
  =\frac{1}{2}\left[-\frac{\partial f_{AB}}{\partial \phi^i}
    f^{AC}(Q_{C}-{\tilde f}_{CD}P^D){1\over b^2}f^{BE}
    (Q_{E}-{\tilde f}_{EF}P^F) {1\over b^2} + \frac{\partial f_{AB}}
    {\partial \phi^i}\frac{1}{b^4}P^AP^B \right],
\end{equation}

\begin{equation}
  =\frac{1}{2}\left[\frac{\partial f^{AC}}{\partial \phi^i}f_{AB}
    (Q_{C}-{\tilde f}_{CD}P^D){1\over b^2}f^{BE}(Q_{E}-{\tilde f}_{EF}P^F) 
    {1\over b^2} + \frac{\partial f_{AB}}{\partial \phi^i}
    \frac{1}{b^4}P^AP^B \right],
\end{equation}

\begin{equation}
  =\frac{1}{2}\left[\frac{\partial f^{AC}}{\partial \phi^i}
    (Q_{C}-{\tilde f}_{CD}P^D){1\over b^2}(Q_{A}-{\tilde f}_{AF}P^F) 
    {1\over b^2} + \frac{\partial f_{AB}}{\partial \phi^i}
    \frac{1}{b^4}P^AP^B \right],
\end{equation}

and also ($F_{rt}=-F_{tr}$)

\begin{equation}
  \frac{1}{8} \frac{\partial \tilde f_{AB}}{\partial \phi^i} 
  F^A_{\phantom{A}\mu\nu} F^B_{\phantom{B}\rho \sigma} 
  \epsilon^{\mu\nu\rho\sigma}=\frac{\partial \tilde f_{AB}}
  {\partial \phi^i} F^B_{\phantom{A} rt}F^A_{\phantom{A} \theta\phi}=
  -\frac{\partial \tilde f_{AB}}{\partial \phi^i}
  f^{BE}(Q_{E}-{\tilde f}_{EF}P^F) {1\over b^2}\frac{P^A}{b^2}
\end{equation}

Then, using $f_{AB}\partial f^{AB}=-f^{AB}\partial f_{AB}$, (\ref{calcul}) becomes:
\begin{equation}
  \frac{\partial V_{eff}}{\partial \phi^i}=
  \left[\frac{\partial f^{AC}}{\partial \phi^i}(Q_{C}-
    {\tilde f}_{CD}P^D)(Q_{A}-{\tilde f}_{AF}P^F) + 
    \frac{\partial f_{AB}}{\partial \phi^i}P^AP^B \right]-
\end{equation}
\begin{equation}
  - 2\frac{\partial \tilde f_{AB}}{\partial \phi^i}f^{BE}(Q_{E}-
  {\tilde f}_{EF}P^F)P^A
\end{equation}
One can easily check that the effective potential (\ref{defpotgen}) satisfies the previous
equation.
\section{Toda equations}
\label{toda}

We rewrite the equations (\ref{eq:phi_eom_3})-(\ref{eq:v_eom_1}) in a form similar to Toda equations. We define
\begin{equation}
  A=2u_2+\alpha_1u_1\, , \qquad B=2u_2+\alpha_2u_1,
\end{equation}
and we obtain the following equivalent system:
\begin{equation}
  \ddot{A}=\left(\frac{1}{2}\alpha_1^2+2\right)Q_1^2\,e^A+\left(\frac{1}{2}
    \alpha_1\alpha_2+2\right)Q_2^2\,e^B\, ,
\end{equation}
\begin{equation}
  \ddot{B}=\left(\frac{1}{2}\alpha_1\alpha_2+2\right)Q_1^2\,e^A + 
  \left(\frac{1}{2}\alpha_2^2+2\right)Q_2^2\,e^B\, .
\end{equation}
For $\alpha_1\alpha_2=-4$ the equations decouple and we obtain:
\begin{equation}
  \ddot{A}=\left(\frac{1}{2}\alpha_1^2+2\right)Q_1^2\,e^A=e^{(A+a)}=
  e^{\bar{A}}\, ,
\end{equation}
\begin{equation}
  \ddot{B}=\left(\frac{1}{2}\alpha_2^2+2\right)Q_2^2\,e^B=e^{(B+b)}=e^{\bar{B}}\, .
\end{equation}
A solution of the equation $\ddot{X}=e^X$ can be written in the following form:
\begin{equation}
  X=\log\left(\frac{2c^{2}}{\sinh^{2}(c(\tau-d))}\right)=
  \log\left(\frac{2c^{2}}{F^2(\tau)}\right)\, .
\end{equation}
The solutions are given by
\begin{equation}
  \bar{A}=A+a=\log\left(\frac{2c_1^{2}}{F_1^2(\tau)}\right)\Rightarrow \,\,\,\,
  A=\log\left(\frac{2c_1^{2}}{F_1^2(\tau)}\frac{2}{Q_1^2(\alpha_1^2+4)}\right)
\end{equation}
\begin{equation}
  \bar{B}=B+b=\log\left(\frac{2c_2^{2}}{F_2^2(\tau)}\right) \Rightarrow \,\,\,\,
  A=\log\left(\frac{2c_2^{2}}{F_2^2(\tau)}\frac{2}{Q_2^2(\alpha_2^2+4)}\right)
\end{equation}

\section{Finding solutions}
\label{fsolutions}
We construct the solution I with $\gamma=1\Leftrightarrow\alpha_{1}\alpha_{2}=-4$.
Without loss of generality we assume $\alpha_{2}<0$.  We obtain the following expression
for the dilaton, $a^2$, and $\tau$ (we define
$\lambda=\frac{\alpha_1}{\alpha_1-\alpha_2}$):
\begin{equation}
  e^{(\alpha_1-\alpha_2)\phi}=e^{A-B}=
  \left(\frac{F_2c_1}{F_1c_2}\right)^2e^{(b-a)}=
  \left(\frac{F_2c_1}{F_1c_2}\right)^2\frac{Q_2^2(\alpha_2^2+4)}
  {Q_1^2(\alpha_1^2+4)}
\end{equation}

\begin{equation}
  a^2=\frac{4}{(\alpha_1^2+4)^{1-\lambda}\,(\alpha_2^2+4)^{\lambda}}
  \left(\frac{F_1Q_1}{c_1}\right)^{2(\lambda-1)}
  \left(\frac{c_2}{F_2Q_2}\right)^{2\lambda}
\end{equation}

\begin{equation}
  \tau=\int\frac{dr}{a^{2}b^{2}}=\frac{1}{(r_{+}-r_{-})}
  \log\left(\frac{r-r_{+}}{r-r_{-}}\right)
\end{equation}

\subsection{Boundary Conditions}\label{sec:bc}

\subsubsection*{$\bullet$ Horizon ($r\rightarrow r_+$, $\tau\rightarrow-\infty$)}

As $r\rightarrow r_{+}$(ie. $\tau\rightarrow-\infty$) the scalar field goes
like \begin{equation} e^{(\alpha_{1}-\alpha_{2})\phi}\sim
  e^{2(c_{1}-c_{2})\tau}\end{equation} so, for $\phi$ finite at the horizon, we require
$c:=c_{1}=c_{2}$.  Also at the horizon \begin{equation} b^{2}\sim(r-r_{+})/a^{2}\sim
  (r-r_+) \left( \frac{r-r_-}{r-r_+} \right)^{\frac{2c}{r_+-r_-}}
  \label{eq:b_at_horiz}
\end{equation}
which necessitates \begin{equation} (r_{+}-r_{-})=2c\end{equation}
\subsubsection*{$\bullet$ Asymptotic infinity ($r\rightarrow\infty$, $\tau\rightarrow0$)}

At infinity, the scalar field tends to 
\[
e^{\left(\alpha_{1}-\alpha_{2}\right)\phi_{\infty}}=
e^{\left(\alpha_{1}-\alpha_{2}\right)\phi(\tau=0)}=
\frac{\alpha_2^2+4}{\alpha^2_1+4}\left(\frac{Q_{2}\sinh(cd_{2})}
  {Q_{1}\sinh(cd_{1})}\right)^{2}\]
which we can write \begin{equation} (\alpha_1^2+4)\bar{Q}_{1}^{2}\sinh^{2}(cd_{1})=
  (\alpha_2^2+4)\bar{Q}_{2}^{2}\sinh^{2}(cd_{2})\label{eq:bc1}
\end{equation}
where\[ \bar{Q}_{i}^{2}=e^{\alpha_{i}\phi_{\infty}}Q_{i}^{2}\]
We also have \begin{eqnarray} a^{2}|_{\tau=0}=1 & = &
  \frac{4}{(\alpha^2_1+4)^{1-\lambda} (\alpha_2^2+4)^{\lambda}} c^{2}\left(Q_{1}
    \sinh(cd_{1})\right)^{-2(1-\lambda)}\left(Q_{2}
    \sinh(cd_{2})\right)^{-2\lambda}\label{eq:bc2}\\
  & = & \frac{4}{(\alpha^2_1+4)^{1-\lambda}(\alpha_2^2+4)^{\lambda}}c^{2}
  \left(\bar{Q}_{1}^{2}\sinh^{2}(cd_{1})\right)^{-(1-\lambda)}
  \left(\bar{Q}_{2}^{2}\sinh^{2}(cd_{2})\right)^{-\lambda}
\end{eqnarray}
Together with (\ref{eq:bc1}) this implies\begin{eqnarray*}
  \sinh(cd_{1}) & = & 2c(\alpha_1^2+4)^{-\frac{1}{2}}\,\bar{Q}_{1}^{-1}\\
  \sinh(cd_{2}) & = & 2c(\alpha_2^2+4)^{-\frac{1}{2}}\, \bar{Q}_{2}^{-1}\end{eqnarray*}

The scalar `charge'
is\[ -\dot{\phi}(\tau=0)=\Sigma=-\frac{2c(\coth(cd_{1})-\coth(cd_{2}))}
{\alpha_{1}-\alpha_{2}}\]
and the mass is \begin{eqnarray*} \dot{a}|_{\tau=0}=M & = &
  \overbrace{\left(\frac{2c\left(Q_{1}
        \sinh(cd_{1})\right)^{-(1-\lambda)}\left(Q_{2}\sinh(cd_{2})
      \right)^{-\lambda}}{[(\alpha^2_1+4)^{1-\lambda}
      (\alpha_2^2+4)^{\lambda}]^{\frac{1}{2}}}\right)}^{1}
  c(\lambda\coth(cd_{2})+(1-\lambda)\coth(cd_{1}))\\
  & = & c(\lambda\coth(cd_{2})+(1-\lambda)\coth(cd_{1})).\end{eqnarray*} The entropy is
given by
\begin{equation}
  \label{misto1}
  S=\pi b^{2}(\tau=-\infty)=\frac{\pi}{4}(\alpha_1^2+4)^{(1-\lambda)}
  (\alpha_2^2+4)^\lambda \left(Q_{1}^{2}e^{2cd_{1}}\right)^{(1-\lambda)}
  \left(Q_{2}^{2}e^{2cd_{2}}\right)^{\lambda}\end{equation}
while the temperature is 
\begin{eqnarray*}
  T & = & \frac{aa'}{2\pi}(\tau=-\infty)=\frac{c}{2\pi}\frac{4}
  {(\alpha_1^2+4)^{(1-\lambda)}(\alpha_2^2+4)^\lambda} 
  \left(Q_{1}^{2}e^{-2cd_{1}}\right)^{-(1-\lambda)}
  \left(Q_{2}^{2}e^{-2cd_{1}}\right)^{-\lambda},
\end{eqnarray*}
and as a check, we note that,
\[ 
ST=\frac{c}{2}=\frac{1}{4}(r_{+}-r_{-}),
\]
and one can also check that
\begin{equation}
  M^2+\Sigma^2-\bar{Q}_1^2-\bar{Q}_2^2=4S^2T^2=c^2.
\end{equation}

\bibliographystyle{JHEP}
\bibliography{paper}

\end{document}